\documentclass[11pt]{article}
\usepackage[letterpaper, left=1in, top=1in, right=1in, bottom=1in,nohead,includefoot]{geometry}
\usepackage{color,charter,graphicx,natbib,fancyhdr,bibentry,enumitem,eso-pic,here,setspace,multirow,amsmath,amssymb,amsfonts,mathdots,sidecap,subcaption}
\usepackage[svgnames,x11names]{xcolor}
\usepackage[pdftex]{hyperref}
\hypersetup{
colorlinks,%
linkcolor=MidnightBlue,  
urlcolor=DarkSlateGray,   
citecolor=RoyalBlue2  
}
\def\blu{\color{RoyalBlue4}}      

\usepackage[color,matrix,arrow,pdf]{xy}\xyoption{all} \xyoption{poly} \xyoption{knot}
\newcommand{\figdir}{.}
\newcommand{\bx}{\mathbf{x}}
\newcommand{\by}{\mathbf{y}}

\newcommand{\bF}{\mathbf{F}}
\newcommand{\bA}{\mathbf{A}}

\newcommand{\bG}{\mathbf{G}}

\newcommand{\bI}{\mathbf{I}} 
\newcommand{\bzero}{\mathbf{0}}

\newcommand{\bphi}{\boldsymbol{\phi}}
\newcommand{\balpha}{\boldsymbol{\alpha}}
\newcommand{\bbeta}{\boldsymbol{\beta}}
\newcommand{\btheta}{\boldsymbol{\theta}}
\newcommand{\bmu}{\boldsymbol{\mu}}
\newcommand{\bnu}{\boldsymbol{\nu}}
\newcommand{\bgamma}{\boldsymbol{\gamma}}
\newcommand{\bomega}{\boldsymbol{\omega}}
\newcommand{\bTheta}{\mathbf{\Theta}}
\newcommand{\bGamma}{\mathbf{\Gamma}}
\newcommand{\bLambda}{\mathbf{\Lambda}}
\newcommand{\bSigma}{\mathbf{\Sigma}}
\newcommand{\bOmega}{\mathbf{\Omega}}

\newcommand{\cD}{\mathcal{D}}
\newcommand{\cG}{\mathcal{G}}
\newcommand{\cM}{\mathcal{M}}

\newcommand{\diag}{\textrm{diag}}
\newcommand{\seq}[2]{#1\,{:}\,#2}

    \usepackage{titlesec} 
\titleformat*{\section}{\normalfont\Large\bfseries\blu }
\titleformat*{\subsection}{\normalfont\large\bfseries\blu }
\titleformat*{\subsubsection}{\normalfont\normalsize\bfseries\blu }
\titleformat*{\paragraph}{\normalfont\normalsize\bfseries\blu }
\titleformat*{\subparagraph}{\normalfont\normalsize\bfseries\blu }

 \def\today{June 2019}

\begin{document}

\setcounter{page}{0}\thispagestyle{empty}

\begin{center} 
{\LARGE\bf\blu Bayesian Forecasting of Multivariate Time Series: 
\smallskip

Scalability, Structure Uncertainty and Decisions}
  
\bigskip\bigskip
{\Large Mike West}\footnote{I was honored to be invited by the Institute of Statistical Mathematics and the Japan Statistical Society  
to present the 2018 Akaike Memorial Lecture. This paper concerns research featured in that address,  
presented at the Annual Conference of the Japanese Federation of Statistical Science Associations, Tokyo, Japan, on September 10th 2018.
I acknowledge the Akaike Memorial Lecture Award committee and the meeting conveners, and constructive comments of invited discussants Chris Glynn and Jouchi Nakajima. Additional thanks go to past students and collaborators on  topics touched on in this paper, many noted as co-authors in the reference list.  Particular thanks are due to Lindsay Berry, Xi Chen and Lutz Gruber on some recent examples and suggestions.

\bigskip

This is an earlier version of the manuscript later published-- together with the invited discussion and reply to the discussion-- in the \href{https://www.ism.ac.jp/editsec/aism/index.html}{Annals of the Institute of Statistical Mathematics},  with DOI:
\href{https://doi.org/10.1007/s10463-019-00741-3}{10.1007/s10463-019-00741-3}

\bigskip 
Mike West, The Arts \& Sciences Professor of Statistics \& Decision Sciences\\ \phantom{.................}\quad \,
                                             Department of Statistical Science, Duke University, Durham NC 27708-0251, USA.
}

\bigskip\bigskip
\today

\bigskip\bigskip\bigskip\bigskip 

{\large\bf \blu Abstract}\end{center}\bigskip 

I overview  recent research advances in Bayesian state-space modeling of multivariate time series. A main focus is on the \lq\lq decouple/recouple'' concept that enables application of state-space models to increasingly large-scale data, applying to continuous or discrete time series outcomes. The scope includes large-scale dynamic graphical models for forecasting and multivariate volatility analysis in areas such as economics and finance,   multi-scale approaches for forecasting discrete/count time series in areas such as commercial sales and demand forecasting, and dynamic network flow models for areas including internet traffic monitoring.  In applications, explicit forecasting, monitoring and decision goals are paramount and should factor into model assessment and comparison, a perspective that is highlighted.     
 
\medskip

{\em\blu Keywords: } Bayesian forecasting; Bayesian model emulation;  decision-guided model assessment; decouple/recouple; dynamic dependency networks; integer count time series; multi-scale  models;    network flows; simultaneous graphical dynamic models;  time series monitoring

\newpage
 
\section{Introduction \label{sec:introduction}} 
Hirotugu Akaike was a seminal contributor to statistical science in its core conceptual bases, in methodology, and in applications.   I  overview  some recent developments in two  areas in which Akaike was an innovator: statistical time series modeling, and statistical model assessment~\citep[e.g.][]{Akaike1974,Akaike1978,Akaike1969BiometrikaAIC,Akaike1981,Akaikepapersbook2012}.
These continue to be challenging areas in basic statistical research as well as in expanding applications. I highlight   recent developments that address statistical and computational scalability of multivariate dynamic models,  and questions of evaluating and comparing models in the contexts of explicit forecasting and decision goals. The content is selective, focused on Bayesian methodology emerging in response to challenges in core and growing areas of time series applications. 

Several classes of models are noted.  In each, advances have used variants of the \lq\lq decouple/recouple'' concept to:  (a) define  flexible dynamic models for individual, univariate series;  (b) ensure  flexibility and relevance of cross-series structures to define coherent multivariate dynamic models; (c)  maximally exploit simple, analytic computations for sequential model fitting (forward filtering) and forecasting; and (d)  enable scalability of resulting algorithms and computations for model fitting, forecasting and use.   Model classes   include 
 dynamic dependency network models~(Section~\ref{sec:DDNMs}), and the more general   simultaneous dynamic graphical models~(Section~\ref{sec:SGDLMs}). These define   flexibility and scalability for conditionally linear dynamic models and address, in particular, concerns for improved multivariate volatility modeling.  Further classes of models   are scalable, structured multivariate and multi-scale approaches for forecasting discrete/count time series~(Section~\ref{sec:DBCMs}), and new classes of dynamic models for complicated and interacting flows of traffic in networks of various kinds~(Section~\ref{sec:DYNETs}).  In each of these areas of recent modeling innovation,  specific problems defining applied motivation are noted. These include problems of time series monitoring in areas including studies of dynamic flows on internet networks,  problems of forecasting with decision goals such as in commercial sales and macroeconomic policy contexts, and problems of financial time series forecasting for portfolio decisions.      
 
Following discussion of background and multivariate Bayesian time series literature in Section~\ref{sec:background},  Sections~\ref{sec:DDNMs}--\ref{sec:DYNETs}  each contact one of the noted model classes, with comments on conceptual innovation linked to decouple/recouple strategies to address the challenges of scalability and modeling flexibility. Contact is also made with questions of model comparisons and evaluation in the contexts of specific applications, with the example areas noted representing ranges of applied fields for which the models and methods are increasingly relevant as time series data scales increase.  Each section ends with some comments on open questions, challenges and hints for future research directions linked to the specific models and applied contexts of the section. 

\section{Background and Perspectives \label{sec:background} } 

\subsection{Multivariate Time Series and Dynamic Models \label{subsec:DLMs} }

Multivariate dynamic linear models (DLMs) with conditionally Gaussian structures remain at the heart of many applications~(\citealp[][chap. 16]{WestHarrison1997};~\citealp[][chaps. 8-10]{PradoWest2010};~\citealp{WestAFMSbook2012}).
In such contexts, denote by  $\by_t$ a $q-$vector time series over equally-spaced discrete time $t$ where each element $y_{j,t}$  follows a univariate DLM:  $y_{j,t} = \bF_{j,t}'\btheta_{j,t}  + \nu_{j,t}$ with known dynamic regression vector $\bF_{j,t},$ latent state vector $\btheta_{j,t}$ and 
zero-mean, conditionally normal  observation errors $\nu_{j,t}$  with, generally, time-varying variances. The state vector evolves via a conditionally linear, Gaussian evolution equation $\btheta_{j,t} = \bG_{j,t} \btheta_{j,t-1} + \bomega_{j,t}$ with known transition matrix $\bG_{j,t}$ and zero-mean, Gaussian evolution errors (innovations) $\bomega_{j,t}.$    The usual assumptions include mutual  independence of the error series and their conditional independence on past and current states. Discount factors are standard in structuring variance matrices of the 
evolution errors and dynamics in variances of observation errors, a.k.a.  volatilities.  See Chapters 4 in each of~\citet{WestHarrison1997} and~\citet{PradoWest2010} for complete details.   In general,  $\bF_{j,t}$ may contain constants, predictor variables, lagged values of the time series, and latent factors that are also modeled.  Then, resulting dynamic latent factor models are not amenable to analytic computations; computationally intensive methods including Markov chain Monte Carlo (MCMC) are needed. 
 
Some multivariate models central to applied work just couple together this set of  univariate DLMs.   Consider special cases when $\bF_{j,t}=\bF_t$ and $\bG_{j,t}=\bG_t$ for all $j=\seq1q,$ so the DLMs share  common regression vectors and evolution matrices.  This defines the class of common components, or exchangeable time series 
models~\citep[][chap. 10]{PradoWest2010} with $\by_t = \bF_t'\bTheta_t + \bnu_t$ where $\bTheta_t = [\btheta_{1,t},\ldots,\btheta_{q,t}]$ and where $\bnu_t$ is the $q-$vector of the observation errors.  
The state evolution becomes a matrix system for $\bTheta_t$ with conditional matrix-normal structure.  Special cases include traditional time-varying vector autoregressions (TV-VAR) when $\bF_t$ includes lagged values of the $y_{j,t}$~\citep[][chap. 9]{Kitagawa1996,PradoWest2010}.
A critical feature is these models allow coupling via a volatility matrix  $V(\bnu_t) = \bSigma_t$ to represent dynamics in cross-series relationships through a role in the matrix-normal  evolution of $\bTheta_t$ as well as in individual volatilities.  
The standard multivariate discount volatility model underlies the class of dynamic inverse Wishart models for $\bSigma_t$, akin to random walks on the implied precision matrices $\bOmega_t = \bSigma_t^{-1}.$  
 Importantly, the resulting analysis for forward filtering and forecasting is easy.   Prior and posterior distributions for $\bSigma_t$ as it changes over time are inverse Wishart,   enabling efficient sequential analysis, and retrospective analysis exploits this for simple posterior sampling over historical periods. 
 Alternative multivariate volatility  models-- such as various multivariate GARCH and others-- are, in contrast often difficult to interpret and challenging to fit, and obviate analytic sequential learning and analysis. 
Though the dynamic Wishart/common components model comes with constraints (noted below) it remains a central workhorse model for monitoring, adapting to and-- in short-term forecasting-- exploiting time-variation in multivariate relationships in relatively low-dimensional series. 

\begin{figure}[t!]
\centering
\includegraphics[width=.9\textwidth]{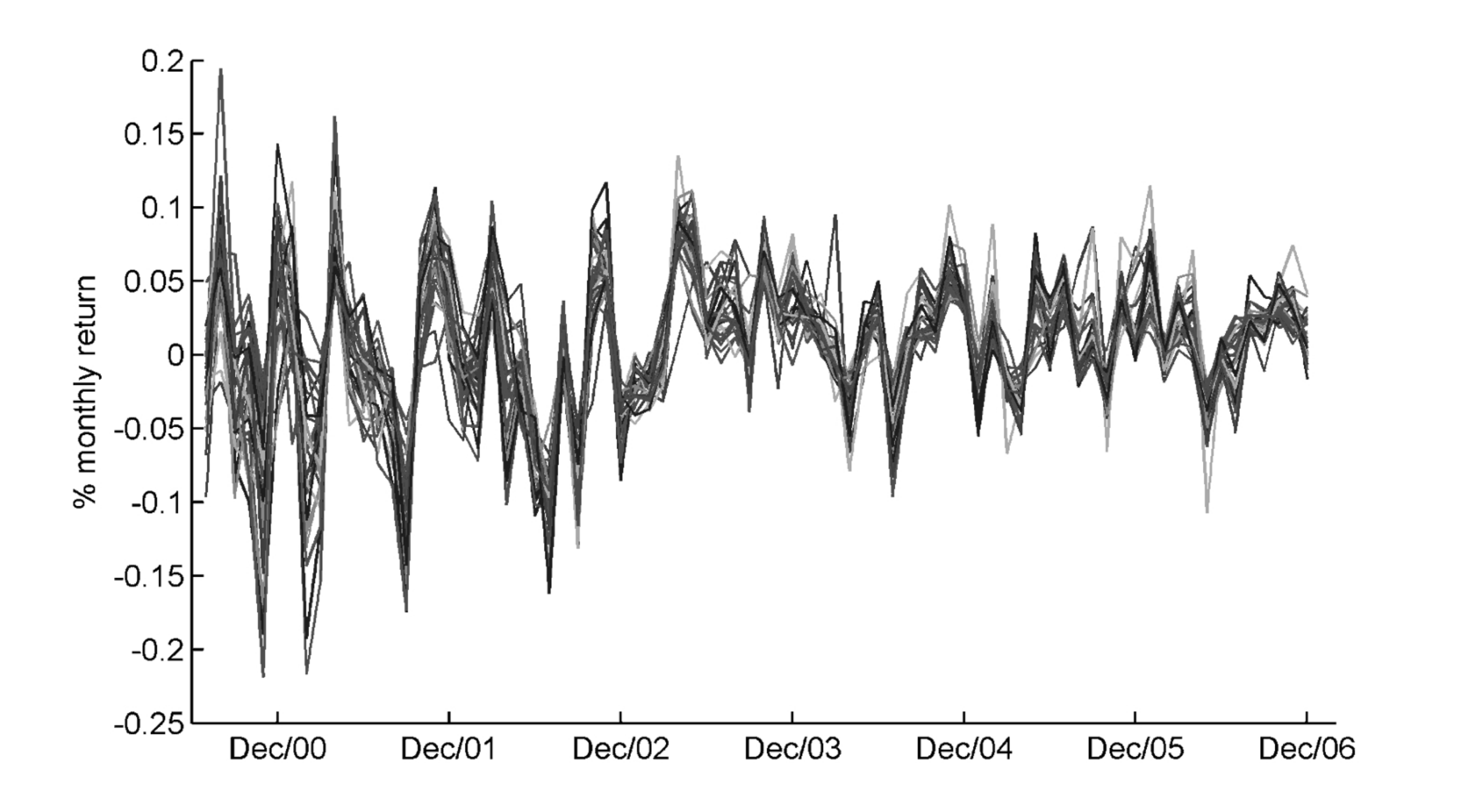}
\caption{Time series of monthly \% financial returns on a set of $q=30$ Vanguard mutual funds over a period of years indicated.  The series represent 18 actively managed funds and 12 index funds that are, in principle, less expensive for an investor.  High dependence across returns series is clear, suggesting that model parameter dimension reduction-- such as offered by graphical model structuring of precision matrices $\bOmega_t$-- is worth exploring. \label{fig:Vanguardreturns}   }    
\end{figure}

\begin{SCfigure}[][t!]
\includegraphics[width=0.55\textwidth]{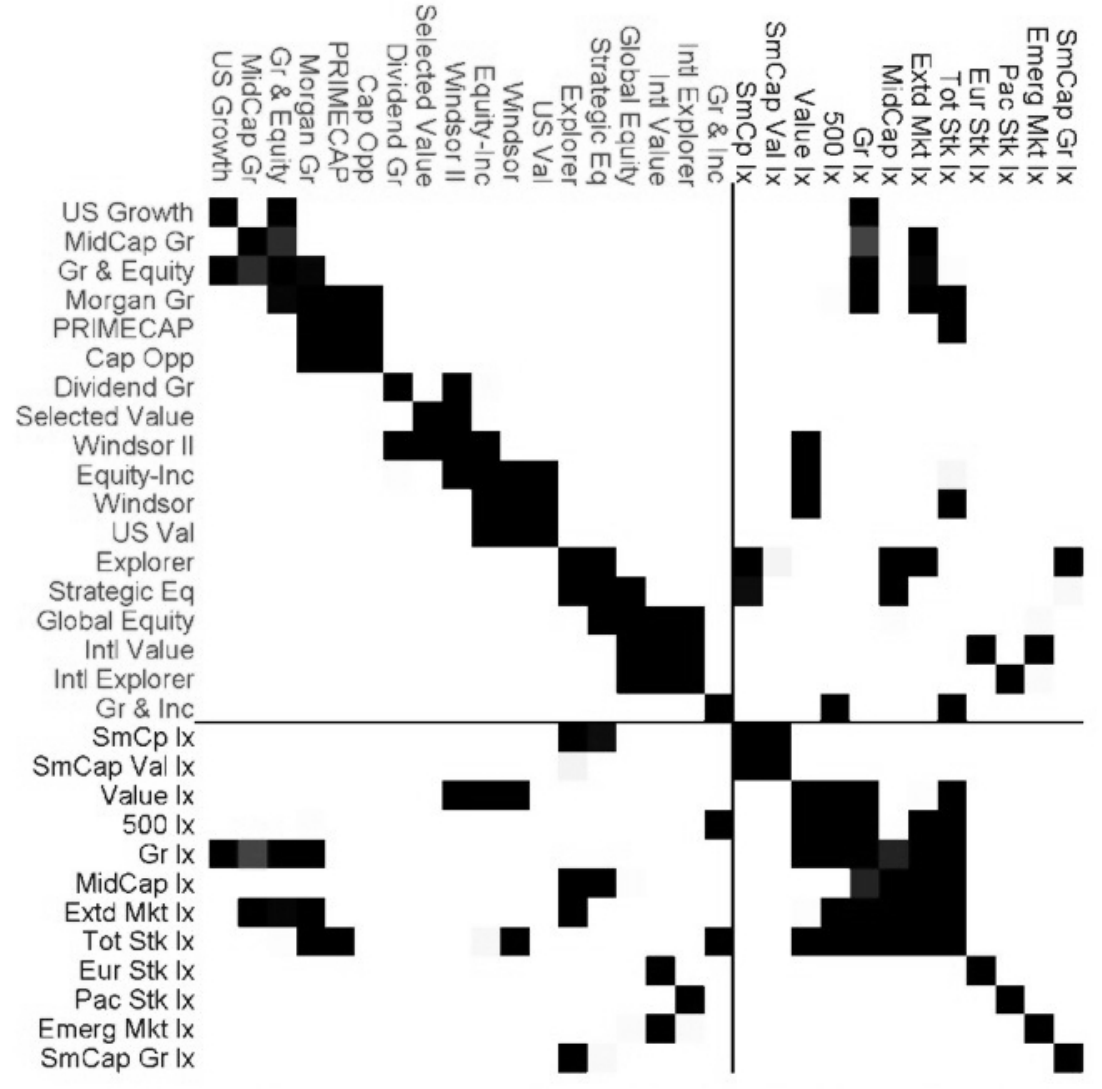}
\caption{Image of posterior probabilities of pairwise edge inclusion in the adjacency matrix of the graph underlying the dynamic precision structure of a multivariate volatility model for 30 monthly Vanguard fund return times series.   The scale runs from 0 (white) to 1 (black) with increasing intermediate grey shades. The horizontal and vertical lines separate the funds into the set of 18 managed funds (above/left) and index funds (below/right). Funds are ordered within each category so that most of the high probability edges cluster near the diagonal.  The figure indicates very concentrated posterior probabilities with multiple edges clearly in and many others excluded, and a strong level of sparsity. \label{fig:Vanguardplotprobabilities}}       
\end{SCfigure}

\subsection{Parameter Sparsity and Dynamic Graphical Model Structuring} 

Interest develops in scaling to higher dimensions $q$, particularly in areas such as financial time series. 
A main concern with multivariate volatility models is/was that of  over-parametrisation of variance matrices $\bSigma_t = \bOmega_t^{-1}$, whether time-varying or not. One natural development to address this was the adaptation of ideas of Bayesian graphical modeling~\citep{Jones2005a,Jones2005}.  A conditional normal model in which each $\bOmega_t$ has zeros in some off-diagonal elements reflects conditional independence structures among the series visualized in an undirected graph: pairs of variables (nodes) are conditionally dependent given all other variables if, and only if, they have edges between them in the graph. The binary adjacency matrix of the graph is a visual of this. Consider the $q=30$ series of monthly returns on a set of Vanguard mutual funds in Figure~\ref{fig:Vanguardreturns} as an example; viewing the image in Figure~\ref{fig:Vanguardplotprobabilities}  as if it were purely white/black, it represents the adjacency matrix of a graph corresponding to off-diagonal zeros in  $\bOmega_t$. The image indicates strong sparsity representing a  small number of non-zero elements; this means significant conditional independence structure and constraints leading to parameter dimension reduction in  $\bSigma_t.$ 

Advances in dynamic modeling that extend the theory of hyper-inverse Wishart distributions for (decomposable) graphical models~\citep{Jones2005a} to common components dynamic models represented the first practical use of graphical models for model parameter dimension reduction.  One of the key features of such extensions is that the analytically tractable forward filtering, forecasting and retrospective posterior sampling methodology is maintained for these models conditional on any specified set of conditional independence relationships, i.e., on any specified graph $\cG$~\citep{Carvalho2007b,Carvalho2007a,Carvalho2007}.    Examples in these papers prove the principle and highlight practical advances in methodology. First, sparsity is often supported by time series data, and forecast accuracy is often improved as a result when using graphs $\cG$ that are sparse and that the data supports. Second, decisions based on data-relevant sparse models are often superior-- in terms of realized outcomes-- to those of the over-parametrized traditional full models, i.e., models with a complete graph and no zeros in $\bOmega_t.$ The statistical intuition that complicated patterns of covariances across series-- and their changes over time-- can be parsimoniously represented with  often far fewer parameters than the full model allows is repeatedly borne out in empirical studies in financial portfolio analyses, econometric and other applications~\citep[e.g.][]{Carvalho2007a,Reeson2009,Wang2009,HaoWang2010,HaoWang2011}. More recent extensions-- that integrate these models into larger Bayesian analyses with MCMC-based variable selection ideas and others~\citep[e.g.][]{Ahelegbey2016,Ahelegbey2016a,Bianchi2019}-- continue to show the benefits of sparse dynamic graphical model structuring.

\subsection{Model Evaluation, Comparison, Selection and Combination \label{subsec:IntroModelScoring}} 

Graphically structured extensions of multivariate state-space models come with significant computational challenges 
unless $q$ is rather small.   Since $\cG$ becomes a choice there is a need to evaluate and explore  models indexed by $\cG$. Some of the above references use MCMC methods in which $\cG$ is an effective parameter, but these are simply not attractive beyond rather low dimensions.  As detailed  in~\cite{Jones2005a}, for example, MCMC can be effective in models with $q\sim 20$ or less with decomposable graphical models, but simply infeasible computationally-- in terms of convergence-- as $q$ increases further. The MCMC approach is simply very poorly developed in any serious applied sense in more general, non-decomposable graphical models, to date; examples in~\cite{Jones2005a} showcase the issues arising with MCMC in even low dimensions ($q\sim 15$ or less) for the general case.  One response to these latter issues has been the development of alternative  computational strategies using stochastic search to more swiftly find and evaluate large numbers of models/graphs $\cG.$ The most effective, to date, build on shotgun stochastic search concepts~\citep[e.g.][]{Jones2005a,HansJasa,HansISBA,ScottCarvalho2008,HaoWang2015}. 
This approach uses a defined score to evaluate a specific model based on one graph $\cG,$ and then explore sets of \lq\lq similar'' graphs that differ in terms of a small number of edges in/out. This process is sequentially repeated to move around the space of models/graphs, guided by the model scores, and can exploit parallelization to enable swift exploration of large numbers of more highly scoring graphs.  

Write $\cD_t$  for all observed data at time $t$ and all other information-- including values of all predictors, discount factors, interventions or changes to model structure, future values of exogenous predictors-- relevant to forecasting.  The canonical statistical score of $\cG$ based on data  over  $t=\seq1n$ is the marginal likelihood value 
$p(\by_{\seq1n}|\cG, \cD_0) = \prod_{t=\seq1n} p(\by_t|\cG, \cD_{t-1} )$.  At time $n$, evaluating this score across  graphs $\cG_1,\ldots,\cG_k$ with specified prior probabilities leads-- by Bayes' theorem-- to posterior model probabilities over these $k$ graphs at this time $t.$   With this score,  stochastic search methods evaluate the posterior over graphs conditional on those found in the search. Inferences and predictions can be defined by model averaging across the graphs in the traditional way~(\citealp[][chap. 12]{WestHarrison1997};~\citealp[][chaps. 12]{PradoWest2010}).   The image in Figure~\ref{fig:Vanguardplotprobabilities} shows partial results of this from one analysis of the Vanguard funds. This simple model used constitutes a local level with discount-based volatility on any graph $\cG$ (precisely as in other examples in~\citealp[][sects. 10.4 \& 10.5]{PradoWest2010}).  The figure shows a high level of implied sparsity in $\bOmega_t$ with strong signals about non-zero/zero entries. 

Traditional model scoring via AIC, BIC and  
variants~(\citealp[][and references therein]{Akaike1974,Akaike1978,Akaike1969BiometrikaAIC,Akaike1981,KitagawaAICetcBook2007};~\citealp[][sect. 2.3.4]{PradoWest2010}) 
define approximations to log marginal likelihoods.   As with full Bayesian analysis based on implied model probabilities, these statistical metrics score models  based on $1-$step ahead forecasting accuracy: the overall score from $n$ observations is the product of realized values of $1-$step forecast densities.  This clearly demarks the applied relevance of this score.  If the view is that a specific \lq\lq true'' data generating process is within the span of a set of selected models $\cG_1,\ldots,\cG_k,$  posterior model probabilities will indicate which are \lq\lq nearest'' to the data; for large $n$, they will concentrate on one \lq\lq Kullback-Leibler'' nearest model~\citep[][sect. 12.2]{WestHarrison1997}.   This is relevant in contexts where the graphical structure is regarded as of inherent interest and one goal is to identify data-supported graphs\citep[e.g.][and references therein]{Tanketal2015}. 

However, more often than not in applications, the role of $\cG$ is as a nuisance parameter and a route to potentially improve accuracy and robustness in forecasting and resulting decisions.   That posterior model probabilities ultimately degenerate is a negative in many contexts, and is contrary to the state-space perspective that changes are expected over time-- changes in relevant model structures as well as state vectors and volatility matrices within any model structure.    Further,   models scoring highly in $1-$step forecasting may be poor for longer-term forecasting and  decisions reliant on forecasts.   While these points have been recognized in recent literature, formal adoption of model evaluation based on other metrics is not yet mainstream.   In an extended class of multivariate dynamic models,~\cite{NakajimaWest2013JBES} and~\cite{NakajimaWest2013JFE} focused on  comparing models based on $h-$step ahead forecast accuracy using horizon-specific model scores: time aggregates of evaluated predictive densities  $p(\by_{t+h-1}|\cG, \cD_{t-1})$. It is natural to consider extensions to score full path forecasts over times $\seq t{t+h-1}$ based on time aggregates of $p(\by_{\seq t{t+h-1}}|\cG, \cD_{t-1})$~\citep{LavineLindonWest2019avs}.  Similar ideas underlie model comparisons for multi-step forecasting in different contexts in~\cite{McAlinnWest2017bpsJOE} and~\cite{McAlinnEtAl2017}, where models rebuilt for specific forecast horizons are shown to be superior to using one model for all horizons, whatever the model selection/assessment method.

Extending this point,   models scoring  highly on statistical metrics may or may not be optimal for specific decisions reliant on forecasts.  While it is typical to proceed this traditional way, increasing attention is needed on decision-guided model selection.  An empirical example in sequential portfolio analysis of the Vanguard mutual fund series highlights this. Using the same model as underlies the statistical summaries on sparse structure in $\bOmega_t$ in 
Figure~\ref{fig:Vanguardplotprobabilities},  stochastic search analysis over graphs $\cG$ was rerun guided by a portfolio metric rather than the conditional posterior model probabilities. For one standard target portfolio loss function, portfolios were optimized and returns realized over the time period, and the score used is simply the overall realized return.  Comparing sets of high probability models with sets of high portfolio return models leads to general findings  consistent with expectations. Models with higher posterior probability are sparse, with typically 20-30\% of edges representing non-zero off-diagonal (and of course time-varying) precision elements; these models tend to generate ranges of realized returns at low to medium portfolio risk levels. Models with higher realized returns are also sparse and generally somewhat sparser, and some of the highest return models have rather low risk.  

\begin{SCfigure}[][b!]
\includegraphics[width=0.55\textwidth]{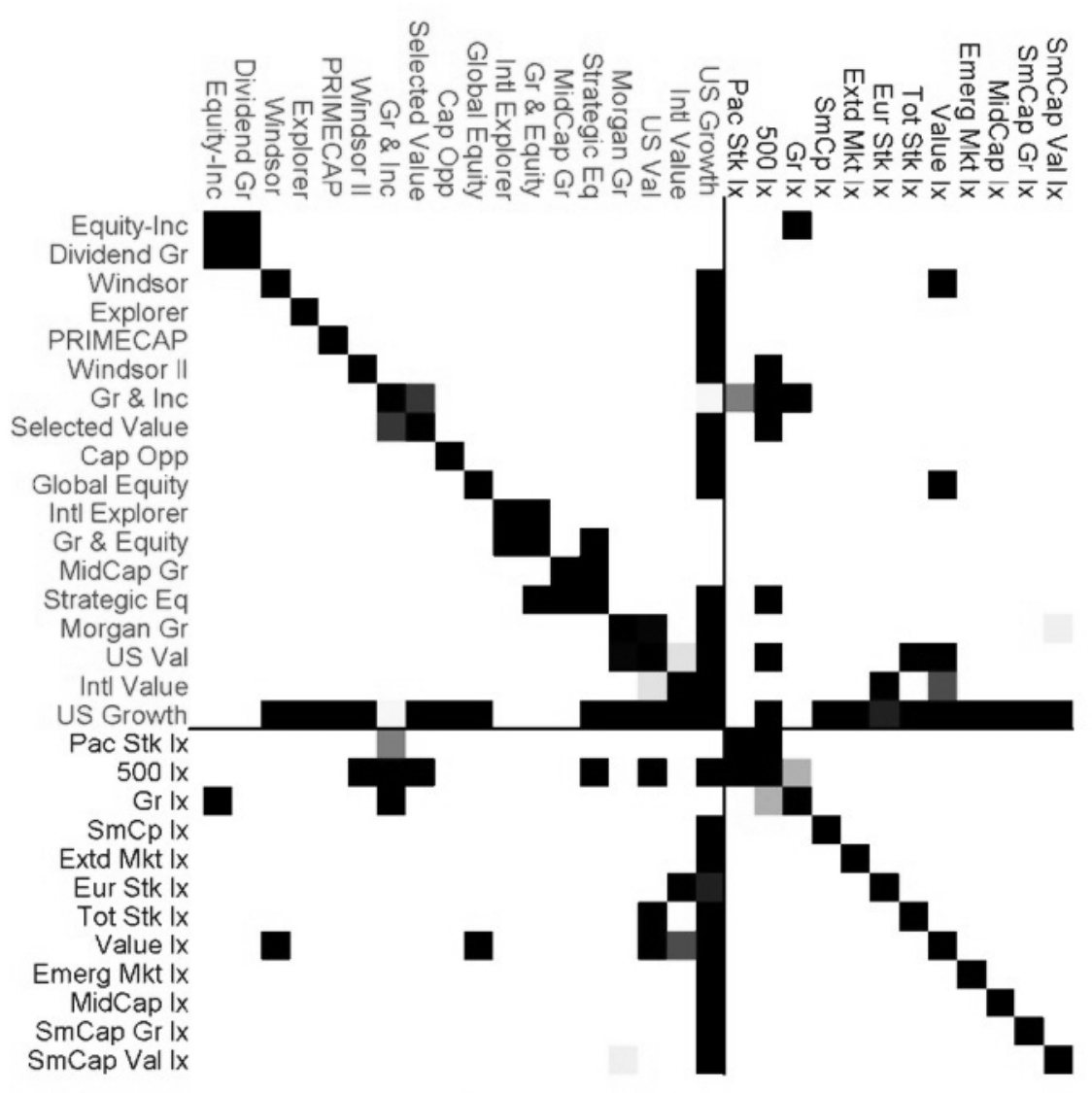}
\caption{Image formatted as in Figure~\ref{fig:Vanguardplotprobabilities}. Now the 
0-1 (white-grey-black) scale  indicates frequency of pairwise edge inclusion across 1{,}000 graphical models identified in stochastic search over graphs guided by a chosen portfolio allocation decision analysis.  These 1{,}000 graphs were those-- out of many millions evaluated-- generating the highest returns over a test time period. Funds are reordered within each of the two categories so that most of the high probability edges cluster near the diagonal.  The figure indicates somewhat different structure and a higher level of sparsity than that in Figure~\ref{fig:Vanguardplotprobabilities}. \label{fig:Vanguardplotportfolios}    }   
\end{SCfigure}

Figure~\ref{fig:Vanguardplotportfolios} shows relative frequencies of edge inclusion across a large number of high scoring portfolio graphs. This appears sparser than in Figure~\ref{fig:Vanguardplotprobabilities}, and has a distinct feature in that one series (US Growth listed last in the first group of managed funds) has a large number of edges to other funds; this series is a \lq\lq hub'' in these top-scoring graphs.  Model search on posterior model probabilities  identifies  graphs that represent the complex patterns of collinearities among the series over time in different ways. Typically, \lq\lq small'' dependencies can be represented in multiple ways, hence the posterior over graphs will tend to identify more candidate edges for inclusion. In contrast, the portfolio decision-guided analysis finds value in sparser graphs with this hub-like structure that is able to generate even weak dependencies among funds other than the hub fund via the one-degree of separation feature. One notable result is that the conditional dependence structure among the index funds (lower right in the figures) appears much sparser under decision-guided analysis than under statistical analysis. Across top graphs in terms of portfolios, the US Growth hub fund is a dominant parental predictor for index funds; Figure~\ref{fig:Vanguardplotportfolios} shows that the set of index funds are rendered almost completely mutually independent conditional on the US Growth fund. This is quite different to the structure across most highly probably models exhibited in Figure~\ref{fig:Vanguardplotprobabilities}.

While in this applied context the structure of relevant graphs is not of primary interest compared to finding good models for portfolio outcomes, this rationalization of differences is illuminating. The example underscores the  point  that different forecasting and/or decision goals should play central roles in model evaluation and comparison.  This is a general point, not restricted to time series and forecasting, but practically  central in such contexts. 
  
\subsection{Challenges and Opportunities} 

Graphical modeling to introduce sparsity-- hence parsimony and potential improved forecasting and decisions--sees increased use in time series as referenced earlier.  However, several issues in existing model classes limit modeling flexibility and scalability.  With studies in 
10s to several 100s of series in areas of finance and macroeconomics becoming routine, some specific issues are noted. 

Common components models-- including the key class of models with TV-VAR components-- are constrained by the common $\bF_t,\bG_t$ structure and hence increasingly inflexible in higher dimensions.  Then, parameter dimension is a challenge. A TV-VAR($p$) component implies $\bF_t$ includes $pq$ lagged values $\by_{\seq{t-1}{t-p}}$, indicating the issue.   Dimension is a key issue with respect to the use of hyper-inverse Wishart (and Wishart) models, due to their inherent inflexibility beyond low dimensions.  The single degree-of-freedom parameter of such models applies to all elements of the volatility matrix, obviating customization of practical importance. 
Larger values of $q$ make search over graphical models increasingly computationally challenging.    

Some of these problems are addressed using more complex models with MCMC and related methods for model fitting.   Models   with dynamic latent factors,  Bayesian model selection priors for elements of state vectors (e.g., subset TV-VAR components), and  involving \lq\lq dynamic sparsity'' are examples
\citep[e.g.][and many others]{Aguilar1999,Aguilar2000,prado:molina:huerta:2006,LopesCarvalho07,DelNegro2008,KoopKorobilis10,Carvalho11,Koop2013,NakajimaWest2013JBES,NakajimaWest2013JFE,ZhouNakajimaWest2014IJF,NakajimaWest2015DSP,Ahelegbey2016,Ahelegbey2016a,kastner2017,NakajimaWest2017BJPS, Bianchi2019,McAlinnWest2017bpsJOE,McAlinnEtAl2017}.
However, one of our earlier noted desiderata is to enable scaling and modeling flexibility in a sequential analysis format, which  conflicts with increasingly large-scale MCMC methods: such methods are often inherently challenging to tune and run, and application in a sequential context requires repeat MCMC analysis each time point.


\section{Dynamic Dependence Network Models\label{sec:DDNMs}}
\subsection{Background} 
Dynamic dependence network models (DDNMs) as in~\cite{ZhaoXieWest2016ASMBI} nucleated the concept of decouple/recouple that has since been more broadly developed. 
DDNMs define coherent multivariate dynamic models via coupling of sets of customized univariate DLMs.   While the DDNM terminology is new, the basic ideas and strategy are much older and have their bases in traditional recursive systems of structural (and/or simultaneous) equation models in econometrics~\citep[e.g.][and references therein]{Bodkinetal1991}. At one level, DDNMs  extend this traditional thinking to time-varying parameter/state-space models within the Bayesian framework.  Connecting to more recent literatures,  DDNM structure has a core directed graphical component that links across series at each time $t$ to define an overall multivariate (volatility) model, indirectly generating a full class of dynamic models for $\bOmega_t$ in the above notation.   In core structure, DDNMs thus extend earlier multiregression dynamic models~\citep{Smith93,Queen94,Queen08,Queen13,Costa14}.

Series ordering  means that these are Cholesky-style volatility models~\cite[e.g.][]{Smith2002,Primiceri05,Shinichiroetal2017,Lopes2018}. The resulting triangular system of univariate models can be decoupled for forward filtering, and then recoupled using theory and direct simulation for coherent forecasting and  decisions.   In elaborate extensions of DDNMs to incorporate dynamic latent factors and other components, the utility of has been evidenced in a range of applications~\citep[e.g.][]{NakajimaWest2013JBES,NakajimaWest2013JFE,NakajimaWest2015DSP,NakajimaWest2017BJPS,ZhouNakajimaWest2014IJF,IrieWest2018portfoliosBA}. 

\subsection{DDNM Structure} 

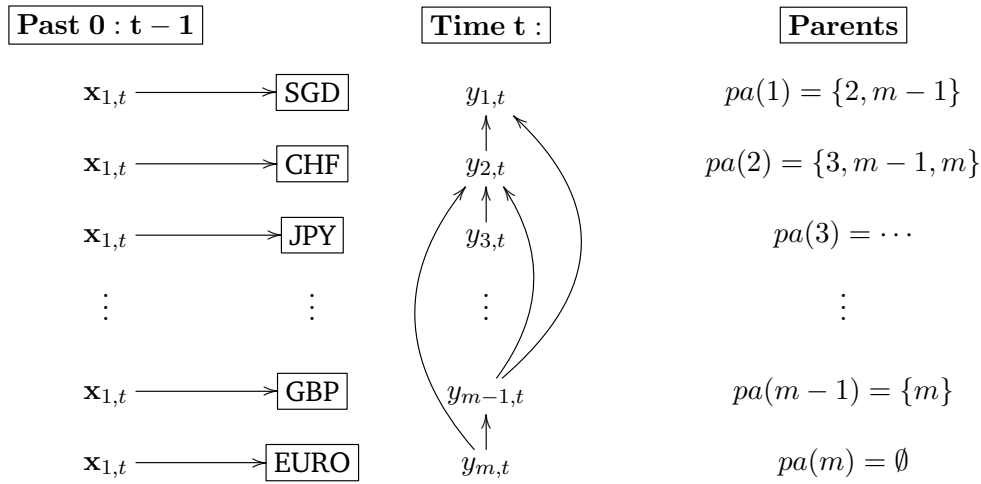
\begin{figure}[b!]
$$\vcenter{\xymatrix@R=.8pc{
 *+[F]{\bf Past\ 0:t-1} 		& & *+[F]{\bf Time\ t:}  			&\,&	*+[F]{\bf Parents}	\\
\bx_{1,t}	\ar[r] 	&       *+[F]{\textrm{SGD}}        & y_{1,t}           &&          pa(1) = \{ 2,m-1 \}     \\
\bx_{1,t}	\ar[r] 	&       *+[F]{\textrm{CHF}}          & y_{2,t}  \ar[u]  &&  pa(2) = \{ 3,m-1,m \}  \\
\bx_{1,t}	\ar[r] 	&        *+[F]{\textrm{JPY}}         & y_{3,t} \ar[u]  && pa(3) = \cdots \\
\vdots	      &        \vdots     	  & \vdots  & &\vdots \\
\bx_{1,t}	\ar[r] 	&        *+[F]{\textrm{GBP}}        & y_{m-1,t}  \ar@/_1.5pc/[uuu]\ar@/_3pc/[uuuu]  && pa(m-1) = \{ m \} \\
\bx_{1,t}	\ar[r] 	&        *+[F]{\textrm{EURO}}        & y_{m,t}\ar[u]\ar@/^2.25pc/[uuuu]   && pa(m)=\emptyset \\
}}
$$
\caption{DDNM for daily prices of international currencies (FX) relative to the US dollar.  In order from top down:   the univariate DLM for the Singapore dollar (SGD) relies on SGD-specific predictors $\bx_{1,t}$ and volatility $\lambda_{1,t},$  along with parental predictors given by the contemporaneous values of prices of the Swiss franc (CHF) and the British pound (GBP); that for the Swiss franc has specific predictors $\bx_{2,t}$ and 
the Japanese yen, British pound and Euro as parents. Further down the list the potential parental predictors are more and more restricted, with the final series $j=q$, here the EURO, having no parents. \label{fig:DDNMschematic} }  
\end{figure}

As in Section~\ref{subsec:DLMs}, take univariate DLMs 
$y_{j,t} = \bF_{j,t}'\btheta_{j,t}  + \nu_{j,t}$ under the usual assumptions. 
In a DDNM,  the regression vectors and state vectors are conformably partitioned  as 
$\bF_{j,t}' = (\bx_{j,t}',\by_{pa(j),t}')$ and $ \btheta_{j,t}'= ( \bphi_{j,t}',  \bgamma_{j,t}')$. 
Here $\bx_{j,t}$ has elements such as constants, predictor variables relevant to series $j,$ lagged values of any of the $q$ series, and so forth; 
$\bphi_{j,t}$ is the corresponding state vector of dynamic coefficients on these predictors.  Choices are customisable to series $j$ and, while each model will tend to have a small number of predictors in $\bF_{j,t},$ there is full flexibility to vary choices across series. Then, $pa(j)\subseteq \{ \seq {j+1}q \}$ is an index set selecting some (typically, a few) of the concurrent values of other series as {\em parental predictors} of $y_{j,t}.$   The series order is important; only series $h$ with $h>j$ can be parental predictors. In graph theoretic terminology,  any series $h \in pa(j)$ is a parent of $j$, while $j$ is a child of $h$. 
The modeling point  is clear: if  I could know the values of other series at future time $t$ I would presumably choose some of them to use to aid in predicting $y_{j,t}$; while this is a theoretical construct, it reduces to a practicable model as noted below.  Then, $\bgamma_{j,t}$ is the state vector of  coefficients on  parental predictors of $\by_{j,t}.$  Third, the random error terms $ \nu_{j,t} $ are assumed independent over $j$ and $t$, with $\nu_{j,t}\sim N(0,1/\lambda_{j,t})$ with   time-varying precision $\lambda_{j,t}.$ 
Figure~\ref{fig:DDNMschematic} gives an illustration of the structure. 

For each series $y_{j,t} = \mu_{j,t} + \by_{pa(j),t}'\bgamma_{j,t} + \nu_{j,t} $ where $\mu_{j,t} =  \bx_{j,t}'\bphi_{j,t}$. 
With $\bmu_t = (\mu_{1,t},\ldots,\mu_{q,t})'$  and $\bnu_t=(\nu_{1,t},\ldots,\nu_{q,t})'$, the multivariate model has structural form
$\by_t = \bmu_t + \bGamma_t\by_t + \bnu_t$ where $\bGamma_t$ is the strict upper triangular matrix with above diagonal rows defined by extending the $\bgamma_{j,t}'$  padded with zeros; that is, row $j$ of $\bGamma_t$ has non-zero elements taken from $\bgamma_{j,t}$ in the columns corresponding to indices in $pa(j).$ With increasing dimension $q,$ models will  involve relatively small  parental sets so that $\bGamma_t$ is sparse. 
The reduced form of the model is $\by_t =\balpha_t +  N(\bA_t\bmu_t, \bSigma_t) $ where $\bA_t = (\bI-\bGamma_t)^{-1}$ so that the mean and precision of $\by_t$ are
\begin{equation} \label{eq:DDNMmeanprecisiongraph} 
\begin{split}
\bA_t \bmu_t & = \bmu_t + \bGamma_t \bmu_t + \bGamma_t^2 \bmu_t + \cdots + \bGamma_t^{q-1} \bmu_t, \\
\bOmega_t & = \bSigma_t^{-1} = (\bI-\bGamma_t)'\bLambda_t (\bI-\bGamma_t) = 
\bLambda_t  - \{ \bGamma_t'\bLambda_t + \bLambda_t\bGamma_t\} + \bGamma_t'\bLambda\bGamma_t
\end{split} 
\end{equation}
where $\bLambda_t = \diag(\lambda_{1,t},\ldots,\lambda_{q,t}).$   
The mean vector $\bA_t \bmu_t$ shows cross-talk through the $\bA_t$ matrix:  series-specific forecast components $\mu_{j,t}$ can have filtered impact on series earlier in the ordering based on parental  sets. In Figure~\ref{fig:DDNMschematic}, series-specific predictions of CHF and GBP impact predictions of SGD directly through the terms from the first row of
$\bGamma_t\bmu_t;$  parental predictors have a first-order effect. Then,  series-specific predictions of EURO also impact predictions of SGD through the $\bGamma_t^2\bmu_t$ term-- EURO is a grandparental predictor of SGD though not a parent. Typically, higher powers of $\bGamma_t$ decay to zero quickly (and $\bGamma_t^q=\bzero$ always) so that higher-order inheritances become negligible; low-order terms can be very practically important.  For the precision matrix $\bOmega_t,$  eqn.~(\ref{eq:DDNMmeanprecisiongraph}) shows first that non-zero off-diagonal elements are contributed by the term $\bGamma_t'\bLambda_t + \bLambda_t\bGamma_t;$   element $\bOmega_{j,h,t}=\bOmega_{h,j,t} \ne 0$ if either $j\in pa(h)$ or $h\in pa(j).$ Second,  the term $\bGamma_t'\bLambda\bGamma_t$ contributes non-zero values to elements $\bOmega_{j,h,t}$ if series $j,h$ are each elements of $pa(k)$ for some other series $k;$ this relates to moralization of directed graphs, adding edges between cases in which $j,h$ are neither parents of the other but share a relationship through common child series in the DDNM. 

\subsection{Filtering and Forecasting: Decouple/Recouple in DDNMs} 

In addition to the ability to customize individual DLMs,  DDNMs allow sequential analysis to be decoupled-- enabling fast, parallel processing-- and  then recoupled for  forecasting and decisions.   The recoupled model gives joint p.d.f.   in compositional form  
$\prod_{j=\seq 1q} p(y_{j,t}|\by_{pa(j),t},\btheta_{j,t},\lambda_{j,t} ,\cD_{t-1})$ which is just the product of normals
  $\prod_{j=\seq 1q} N(y_{j,t}|\bF_{j,t}'\btheta_{j,t},1/\lambda_{j,t})$ where $N(\cdot|\cdot,\cdot)$ is the normal p.d.f.    For sequential updating,  this gives the time $t$ likelihood function for $\btheta_{\seq 1q,t},\lambda_{\seq 1q,t}; $   independent conjugate priors across series are conjugate and lead to independent posteriors. 
Using discount factor DLMs, standard forward-filtering analysis propagates prior and posterior distributions for $(\btheta_{j,t},\lambda_{j,t})$ over time using standard normal/inverse gamma distribution theory~\citep[][chap. 4]{PradoWest2010}  independently across series.  Sequential filtering is analytic and scales linearly in $q.$ 

Forecasting involves recoupling and, due to the roles of parental predictors and that   practicable models often involve lagged elements of $\by_\ast$ in $\bx_{j,t}$, is  effectively accessed via direct simulation. \cite{ZhaoXieWest2016ASMBI} discuss recursive analytic computation of $k-$step ahead mean vectors and variance matrices-- as well as precision matrices-- but full inferences and decisions will often require going beyond these partial and marginal summaries, so simulation is preferred.   The ordered structure of a DDNM means that simulations are performed recursively using the implicit  compositional representation.  At time $t,$ the normal/inverse gamma posterior $p(\btheta_{q,t},\lambda_{q,t}|\cD_t)$ is trivially sampled to generate samples from $p(\btheta_{q,t+1},\lambda_{q,t+1}|\cD_t)$ and then $p(y_{q,t+1}|\cD_t).$  Simulated $y_{q,t+1}$ values are then passed up to the models for other series $j<q$ for which they are required as parental predictors. Moving to series $q-1,$ the process is repeated to generate $y_{q-1,t}$ values and, as a result, samples from $p(y_{\seq{q-1}q,t+1}|\cD_t).$  Recursing leads to full Monte Carlo samples drawn directly from $p(\by_{t+1}|\cD_t).$ Moving to $2-$steps ahead, on each Monte Carlo sampled vector $\by_{t+1}$ this process is repeated with posteriors for DLM states and volatilities conditioned on those values and time index incremented by 1. This results in sampled $\by_{t+2}$ vectors jointly with the conditioning vales at $t+1,$ hence samples from $p(\by_{\seq{t+1}{t+2}}|\cD_t).$  Continue this process to $k-$steps ahead to generate full Monte Carlo samples of the path of the series into the future, i.e., generating from  $p(\by_{\seq{t+1}{t+k}}|\cD_t).$  Importantly, the analysis is as scalable as theoretically possible; the computational burden scales as the product of $q$ and the chosen Monte Carlo sample size, and can  exploit partial parallelisation. 

\subsection{Perspectives on Model Structure Uncertainty} 

Examples in~\cite{ZhaoXieWest2016ASMBI} with $q=13$ financial time series illustrate the analysis, with foci on $1-$step and $5-$step forecasting and resulting portfolio analyses.  There the univariate DLM for series $j$ has a local level and some lagged values of series $j$ only, representing custom time-varying autoregressive (TVAR) predictors for each series. The model specification relies on a number of parameters and hence there are model structure uncertainty questions. 
Write $\cM_j$ for a set of  $|\cM_j|$  candidate models for series $j,$  with elements $\cM_j^r$ indexed by specific models $r\in \{ \seq 1{|\cM_j|}\}.$  In~\cite{ZhaoXieWest2016ASMBI}, 
each $\cM_j^r$ involved one choice of the  TVAR order for series $j,$  one value of each of a set of discount factors (one for each of $\bphi_{j,t}, \bgamma_{j,t}, \lambda_{j,t}$) from a finite grid of values, and one choice of the parental set $pa(j)$ from all possibilities.   Importantly, each of these is series specific and the model evaluation and comparison questions can thus be decoupled and addressed using training data to explore, compare and score models.  Critically for scalability,  decoupling means that this involves a total of $\sum_{j=\seq 1q} |\cM_j|$ models for the full vector series, whereas a direct multivariate analysis would involve a much more substantial set of $\prod_{j=\seq 1q} |\cM_j|$ models; for even relatively small $q$ and practical models, this is a major  computational advance. 

A main interest in~\cite{ZhaoXieWest2016ASMBI} was on forecasting for portfolios, and the benefits of use of DDNMs are illustrated there.  Scoring models on portfolio outcomes is key, but that paper also considers comparisons with traditional Bayesian model scoring via posterior model probabilities.  One interest was to evaluate discount-weighted marginal likelihoods and resulting modified model probabilities that, at each time point, are based implicitly on exponentially down-weighting contributions from past data. This   acts to avoid model probabilities degenerating and  has the flavor of representing stochastic changes over time in model space.  Specifically, a model power discount factor $\alpha\in (0,1]$   modifies the time $n$ marginal likelihood on   $\cM$ to give  log score 
$\sum_{t=\seq 1n} \alpha^{n-t} \log (p(\by_t|\cM, \cD_{t-1})).$ 
In terms of model probabilities at time $t,$ the implication is that 
$Pr(\cM|\cD_t) \propto Pr(\cM|\cD_{t-1})^\alpha p(\by_t|\cM, \cD_{t-1})$, i.e., a modified form of Bayes' theorem that \lq\lq flattens'' the prior probabilities over models using the $\alpha$ power prior to updating via the current marginal likelihood contribution. 
At $\alpha=1$ this is the usual marginal likelihood.   Otherwise, smaller values of $\alpha$  discount history in weighting models currently, and allow for    adaptation over time in model space if the data suggests that different models are more relevant over different periods of time.  Examples in~\cite{ZhaoXieWest2016ASMBI} highlight this; in more volatile periods of time (including the great recessionary years 2008-2010) models with lower discount factors on state vectors and volatilities tend to be preferred for some series, while preference for higher values and, in some cases, higher TVAR order increases in more stable periods. That study also highlights the implications for identifying relevant parental sets for each series and how that changes through time. 

\cite{ZhaoXieWest2016ASMBI} show this modified model weighting can yield major  benefits. Short and longer-term forecast accuracy is generally improved with $\alpha<1,$ but the analysis becomes over-adaptive as $\alpha$ is reduced further. In comparison, portfolio outcomes-- in terms of both realized returns and risk measures-- are significantly improved with $\alpha $ just slightly less than 1-- but clearly lower than 1-- but deteriorate for lower values.  The power discounting idea~\citep{Xie12,ZhaoXieWest2016ASMBI} was used historically in Bayesian forecasting~\citep[][p.445]{WestHarrison1989} and has more recently received attention linking to parallel historical 
literature where discount factors are called \lq\lq forgetting" factors~\citep{Raftery10,Koop2013}.  The basic idea and implementation are simple; 
in terms of a marginal broadening of perspectives on model structure uncertainty and model weighting, this power discounting is a trivial technical step and can yield substantial practical benefits.

\subsection{Challenges and Opportunities} 
    
Scaling DDNMs to increasingly large problems exacerbates the issue of model structure uncertainty. An holistic view  necessitates demanding computation for search over spaces of models.  DDNMs contribute a major advance in reducing the dimension of model space and open the opportunity for methods such as variants of stochastic search 
to be applied in parallel to sets of decoupled univariate DLMs. Nevertheless, scaling to 00s or 000s of series challenges any such approach. 
    
DDNMs require a specified order of the $q$ series.     This is a decision made to structure the model, but is otherwise   typically not 
of primary interest.  It is not typically a choice to be regarded as a \lq\lq parameter''  and, in some applications, should be regarded as part of the substantive specification. For example, with lower-dimensional series in macroeconomic and financial applications, the ordering may reflect economic reasoning and theory, as I (with others) 
have emphasized in related work~\citep[e.g.][]{Primiceri05,NakajimaWest2013JBES,NakajimaWest2013JFE,ZhouNakajimaWest2014IJF}. 

Theoretically, series order is irrelevant to predictions  as they rely only 
on the resulting precision matrices (and regression components) that are order-free. 
Practically, of course, the specification of priors and specific computational methods rely on the chosen ordering 
and so prediction results will vary  under different orders.  There are then questions of 
more formal approaches to defining ordering(s) for evaluation,   and a need to consider approaches to relaxing the requirement for ordering to begin.

\section{Simultaneous Graphical Dynamic Linear Models \label{sec:SGDLMs}} 

\subsection{SGDLM Context and Structure} 

\begin{figure}[b!]
$$\vcenter{\xymatrix@R=.8pc{
 *+[F]{\bf Past\ 0:t-1} 		& &	 *+[F]{\bf Time\ t:}  		& \qquad	&	*+[F]{\bf Parents}	\\
\bx_{1,t}	\ar[r] 	&       *+[F]{\textrm{SGD}}        & y_{1,t} \ar@/_1.85pc/[d]       &    &          sp(1) = \{ 2,m-1 \}     \\
\bx_{1,t}	\ar[r] 	&       *+[F]{\textrm{CHF}}          & y_{2,t}  \ar[u]  &&  sp(2) = \{ 1,3, m \}  \\
\bx_{1,t}	\ar[r] 	&        *+[F]{\textrm{JPY}}         & y_{3,t} \ar[u]\ar@/_0.85pc/[dd]  && sp(3) = \cdots \\
\vdots	      &        \vdots     	  & \vdots  & & \vdots \\
\bx_{1,t}	\ar[r] 	&        *+[F]{\textrm{GBP}}        & y_{m-1,t} \ar@/^1.5pc/[d]   \ar@/_3pc/[uuuu]  && sp(m-1) = \{3,m\} \\
\bx_{1,t}	\ar[r] 	&        *+[F]{\textrm{EURO}}        & y_{m,t}\ar[u]\ar@/^2.25pc/[uuuu]   && sp(m)=\{m-1\} \\
}}
$$
\caption{Schematic of SGDLM for FX time series to compare with the DDNM in Figure~\ref{fig:DDNMschematic}.  \label{fig:SGDLMschematic} }
%
\bigskip\bigskip
\centering
\includegraphics[width=\textwidth]{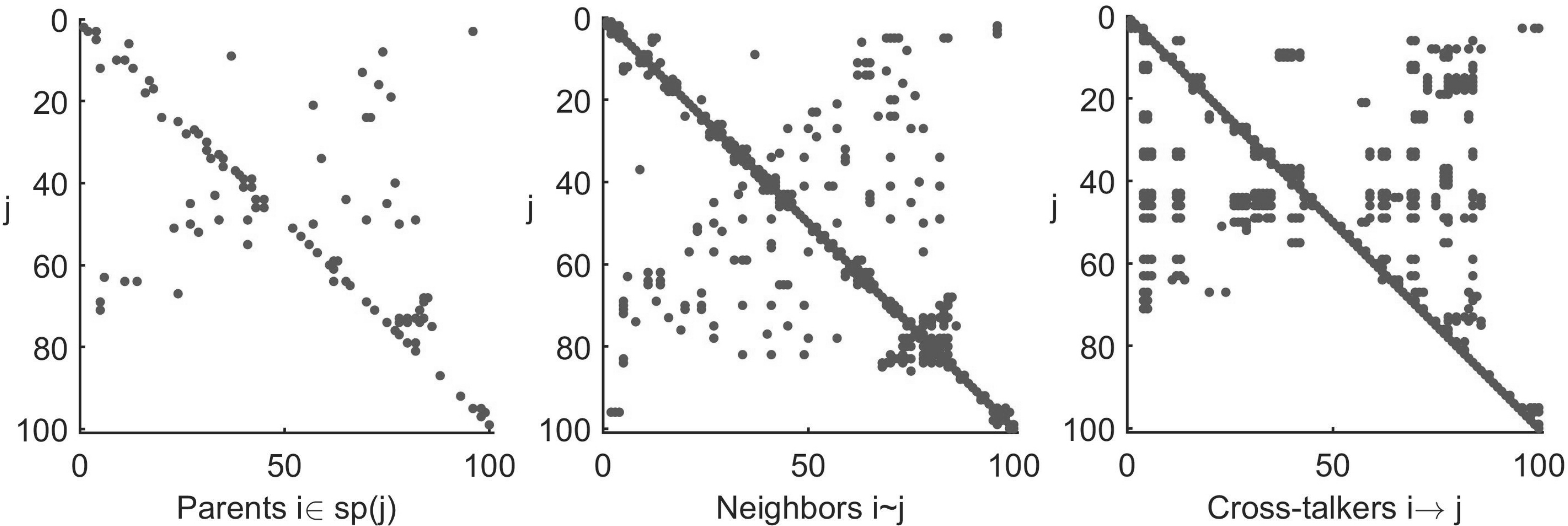}
\caption{{\em Left}: Indicator of simultaneous parents in an example SGDLM with $q=100;$ non-zero elements in each row of $\bGamma_t$ are shaded. 
{\em Center}: Implied non-zero/zero pattern in precision matrix $\bOmega_t.$  {\em Right:} Implied non-zero/zero pattern in prediction cross-talk matrix $\bA_t = (\bI-\bGamma_t)^{-1}$. }
\label{fig:SGDLMspyplots}       
\end{figure}

As introduced  in~\cite{GruberWest2016BA}, 
SGDLMs generalize DDNMs by allowing any series to be a contemporaneous predictor of any other. To reflect this, the parental set for series $j$ is now termed a set of simultaneous parents, denoted by $sp(j) \subseteq \{ \seq 1q\backslash j \},$  with the same DLM model forms, i.e., 
$y_{j,t} = \bF_{j,t}' \btheta_{j,t} + \nu_{j,t} = \bx_{j,t}' \bphi_{j,t} + \by_{sp(j),t}'\bgamma_{j,t} + \nu_{j,t}$ and other assumptions unchanged. 
Figure~\ref{fig:SGDLMschematic} shows an  example to compare with Figure~\ref{fig:DDNMschematic}; directed edges can point down as well as up the list of series-- model structure is series order independent. 
The implied joint distributions are as in DDNMs but now $\bGamma_t$-- while generally still sparse and with diagonal zeros-- does not need to be upper triangular.  
This resolves the main constraint on DDNMs  while leaving the overall structural form of the model unchanged.  DDNMs are special cases when $sp(j)=pa(j)$ and $\bGamma_t$ is upper triangular. The reduced form of the full multivariate model is 
$\by_t =\balpha_t +  N(\bA_t\bmu_t, \bSigma_t) $ with prediction cross-talk matrix $\bA_t = (\bI-\bGamma_t)^{-1}$;  the mean vector and precision matrix are as in 
eqn.~(\ref{eq:DDNMmeanprecisiongraph}),
but now the equation for $\bA_t \bmu_t$ is extended to including sums of terms $ \bGamma_t^k \bmu_t$ for $k\ge q.$  In general, sparse $\bGamma_t$ implies that this infinite series converges as the higher-order terms quickly become negligible. Cross-talk is induced among series as in DDNMs, as is graphical model structure of $\bOmega_t$ in cases of high enough levels of sparsity of parental sets $sp(j)$ and hence of $\bGamma_t$; see Figure~\ref{fig:SGDLMspyplots} for illustration. 

\subsection{Recoupling for Forecasting in SGDLMs} 

\begin{figure}[t!]
$$\vcenter{\xymatrix@R=.8pc{
 *+[F]{\bf Past\ 0:t} 		&	 *+[F]{\bf  Synthetic\ states:}  		& *+[F]{\bf Recouple:}  	&	*+[F]{\bf Synthetic\ futures:}	\\
 *+[F]{\textrm{SGD}} 	\ar[r] 	&   \{ \btheta,\lambda\}_{1,t+1:t+k} \ar@/^2pc/[dddr]    &        &    &           \\
 *+[F]{\textrm{CHF}} 	\ar[r] 	&     \{ \btheta,\lambda\}_{2,t+1:t+k}  \ar@/^1pc/[ddr]            &  &&     \\
*+[F]{\textrm{JPY}}	\ar[r] 	&        \{ \btheta,\lambda\}_{3,t+1:t+k}    \ar [dr]       &   &&  \\
\vdots	      &        \vdots     	  &    *+[F]{ \{ \bA\bmu,\bOmega \}_{t+1:t+k} }\ar@2[r]& \by_{t+1:t+k}  \\ 
 *+[F]{\textrm{GBP}}	\ar[r] 	&       \{ \btheta,\lambda\}_{q-1,t+1:t+k}   \ar[ur]      &     &&   \\ 
  *+[F]{\textrm{EURO}} 	\ar[r] 	&       \{ \btheta,\lambda\}_{q,t+1:t+k}  \ar@/_2pc/[uur]        &    &&   \\
}}
$$
\caption{Decoupled DLM simulations followed by recoupling for forecasting in SGDLMs. \label{fig:SGDLMforecastingschematic} }
\end{figure}
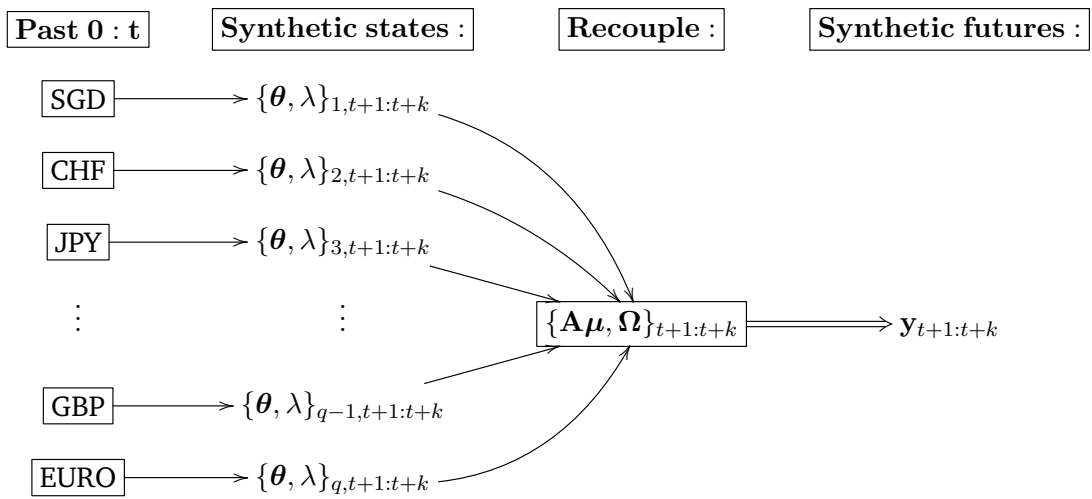

Prediction of future states and volatilities uses simulation in the decoupled  DLMs; these are then recoupled to full joint forecast distributions to simulate the multivariate  outcomes. At time $t-1,$  the SGDLM analysis~\citep{GruberWest2016BA,GruberWest2017ECOSTA} constrains the prior $p(\btheta_{\seq 1q,t},\lambda_{\seq 1q,t} |\cD_{t-1})$ as a product of conjugate normal/inverse gamma forms for the $\{ \btheta_{j,t},\lambda_{j,t} \} $ across series. These are exact in DDNM special cases, and (typically highly) accurate approximations in sparse SGDLMs otherwise. These priors are easily simulated (in parallel) to compute Monte Carlo samples of the implied $\bA_t\bmu_t,\bOmega_t$;  sampling the full $1-$step predictive distribution to generate synthetic $\by_t$ follows trivially.   Each sampled set of states and volatilities underlies conditional sampling of those at the next time point,  hence samples of $\by_{t+1}$.  This process is recursed into the future to generate Monte Carlo of predictive distributions over multi-steps ahead; see Figure~\ref{fig:SGDLMforecastingschematic}.  This involves only direct simulation, so is efficient and scales linearly in $q$ as in simpler DDNMs.

\subsection{Decouple/Recouple for Filtering in SGDLMs} 

The recoupled SGLM no longer defines a compositional representation of the conditional p.d.f. for $\by_t$ given all model quantities (unless $\bGamma_t$ is diagonal). The
 p.d.f. is now of the form given by $|\bI-\bGamma_t|_+\ \prod_{j=\seq 1q} N(y_{j,t}|\bF_{j,t}'\btheta_{j,t},1/\lambda_{j,t})$ where $|\ast|_+$ is the absolute value of the determinant of the matrix argument $\ast$.   Independent normal/inverse gamma priors for the $\{ \btheta_{j,t},\lambda_{j,t} \} $ imply a joint posterior proportional to 
$|\bI-\bGamma_t|_+ \ \prod_{j=\seq 1q} g_j(\btheta_{j,t},\lambda_{j,t}|\cD_t)$  where the $g_j(\cdot|\cdot)$ are the normal/inverse gamma posteriors from each of the decoupled DLMs. 
The $1-$step filtering update is only partly decoupled; the determinant factor recouples across series, involving (only)  state elements related to parental sets.  For sequential filtering to lead to decoupled conjugate forms at the next time point, this posterior must be approximated by a product of normal/inverse gammas.  In practical contexts with larger $q$, the $sp(j)$ will be small sets and so $\bGamma_t$ will be rather sparse; increasing sparsity means that $|\bI-\bGamma_t|_+$ will be closer to 1. Hence the posterior will be almost decoupled and close to a product of conjugate forms. This insight underlies an analysis strategy~\citep{GruberWest2016BA,GruberWest2017ECOSTA} that uses importance sampling for Monte Carlo evaluation of the joint (recoupled) time $t$ posterior, followed by a variational Bayes' mapping to decoupled  conjugate forms. 

\begin{figure}[htbp!]
$$\vcenter{\xymatrix@R=.8pc{
 		&	 *+[F]{\bf  States:}  		& *+[F]{\bf Decouple:}  	&	*+[F]{\bf Observation:}	\\
 *+[F]{\textrm{SGD}} 	 &   \{ \btheta,\lambda\}_{1,t}       &        &    &           \\
 *+[F]{\textrm{CHF}} 	 &     \{ \btheta,\lambda\}_{2,t}             &  &&     \\
*+[F]{\textrm{JPY}}	 	&       \{ \btheta,\lambda\}_{3,t}        &   &&  \\
\vdots	      &        \vdots     	  &    *+[F]{ \{ \btheta,\lambda  \}_{1:q,t} }\ar@/_2pc/[uuul] \ar@/_1pc/[uul] \ar[ul] \ar@/^1pc/[ddl] \ar[dl]        & \by_t \ar@2[l] \\ 
 *+[F]{\textrm{GBP}}	 	&       \{ \btheta,\lambda\}_{q-1,t}     &     &&   \\ 
  *+[F]{\textrm{EURO}}  &       \{ \btheta,\lambda\}_{q,t}          &    &&   \\
}}
$$
\caption{Filtering updates in  SGDLMs. The coupled joint posterior $p(\btheta_{1:q,t},\lambda_{1:q,t}|\cD_t)$ is evaluated by importance sampling, and then decoupled using variational Bayes to define decoupled conjugate form posteriors for the states and volatilities in each univariate model. \label{fig:SGDLMupdateschematic} }
\end{figure}
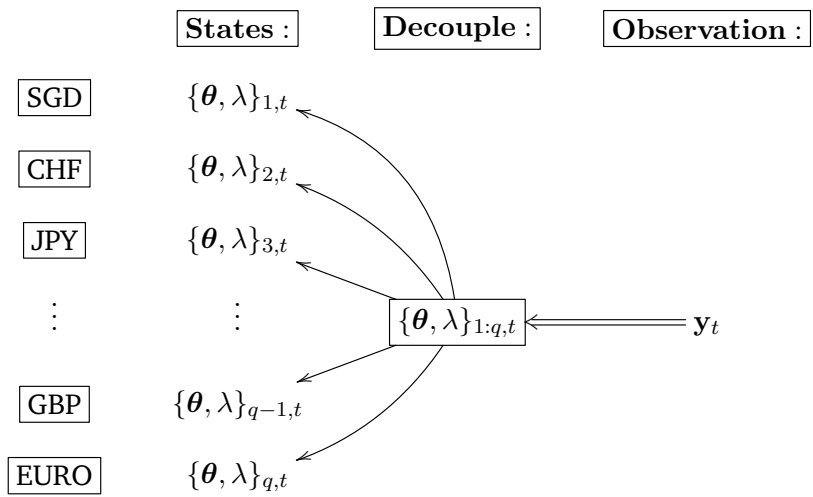

The posterior proportional to  $|\bI-\bGamma_t|_+\ \prod_{j=\seq 1q} g_j(\btheta_{j,t},\lambda_{j,t}|\cD_t)$ defines a perfect context for importance sampling (IS) Monte Carlo when-- as is typical in practice-- the determinant term is expected to be relatively modest in its contribution. Taking the product of the $g_j(\cdot|\cdot)$ terms as the importance sampler yields normalized IS weights proportional to $|\bI-\bGamma_t|_+$ at sampled values of $\bGamma_t.$  In sparse cases, these weights will vary around 1, but tend to be close to 1; in special cases of DDNMs, they are exactly 1 and IS is exact random sampling. Hence posterior inference at time $t$ can be  efficiently based on IS sample and weights, and monitored through standard metrics such as the effective sample size  $ESS = 1/\sum_{i=\seq 1I} w_{i,t}^2$ where $w_{i,t}$ represents the IS weight on each Monte Carlo sample $i=\seq 1I$.  
To complete the time $t$ update and define decoupled conjugate form posteriors across the series requires an approximation step. This is done via a variational Bayes (VB) method that approximates the posterior IS sample by a product of normal/inverse gamma forms-- a mean field approximation-- by minimizing the Kullback-Leibler (KL) divergence of the approximation from the IS-based posterior; see Figure~\ref{fig:SGDLMupdateschematic}. This is a context where the optimization is easily computed and, again in cases of sparse $\bGamma_t,$ will tend to be very effective and only a modest modification of the product of the $g_j(\cdot|\cdot)$ terms. Examples in~\cite{GruberWest2016BA} and \cite{GruberWest2017ECOSTA} bear this out in studies with up to $q=401$ series in financial forecasting and portfolio analysis. 
 
\subsection{Entropy-based Model Assessment and Monitoring} 
Examples referenced above demonstrate  scalability and efficiency of SGDLM analysis
(with parallel implementations--~\citealp{GruberSGDLMcode}) and improvements in forecasting and decisions relative to standard models.   Examples include  $q=401$ series of daily stock prices on companies in the S\&P index along with the index itself. The ability to customize individual DLMs improves characterisation of short-term changes and series-specific volatility, and selection of the $sp(j)$ defines adaptation to dynamics in structure across subsets of series that improves portfolio outcomes across a range of models and portfolio utility functions. 

\begin{figure}[h!]
\centering
\includegraphics[width=0.85\textwidth]{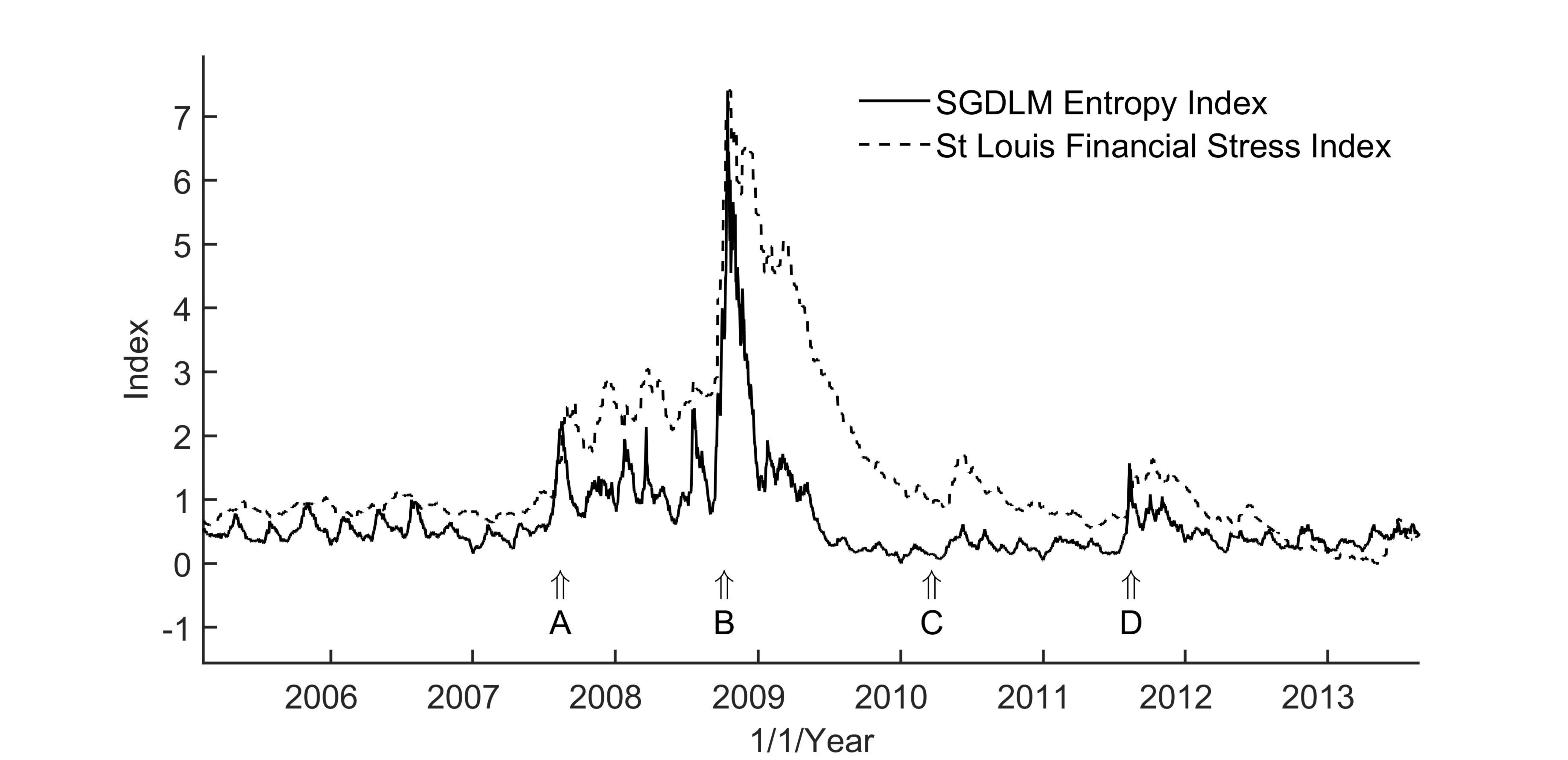} 
\caption{Trajectories of the daily Entropy Index $K_t$ in SGDLM analysis of $q=401$ S\&P series, and the weekly St.~Louis Federal Reserve Bank Financial Stress Index,  over 2005-2013 with 4 key periods indicated.   A:  Aug 2007 events including the UK government intervention on Northern Rock bank,  generating major news related to the subprime loan crisis; B: Oct 2008 US loans \lq\lq buy-back'' events and the National Economic Stimulus Act; C: Mar 2010 initial responses by the European Central Bank
to the ``Eurozone crisis''; D: Aug 2011 US credit downgraded by S\&P.
\label{fig:KLindexSGDLM} }      
\end{figure}

Those examples also highlight sequential monitoring to assess efficacy of the IS/VB analysis. 
At each time $t$ denote by $E_t$ the evaluated ESS for IS recoupling and by $ K_t$ the minimized KL divergence in VB decoupling. These are inversely related:
IS weights closer to uniform lead to high $E_t$ and low $K_t;$  ~\cite{GruberWest2016BA} discuss theoretical relationships and emphasize monitoring.  If a period of low $K_t$ breaks down to higher values, then recent data indicates changes that may be due to increased volatility in some series or changes in cross-series relationships. This calls for intervention to modify the model through changes to current posteriors, discount factors, and/or parental sets.   Simply running the analysis on one model class but with no such intervention~\citep[as in][]{GruberWest2017ECOSTA} gives a benchmark analysis; over a long period of days, the resulting $K_t$ series is shown in Figure~\ref{fig:KLindexSGDLM}.

Figure~\ref{fig:KLindexSGDLM} gives context and comparison with a major financial risk index-- the St.~Louis Federal Reserve Bank Financial Stress Index~\citep{kliesensmith2010}-- widely regarded as local predictor of risk in the global financial systems.   Comparison with the $K_t$ \lq\lq Entropy Index'' is striking.   As a purely statistical index based on stock price data rather than the macroeconomic and FX data of the St.~Loius index,   $K_t$  mirrors the St.~Louis index but shows the ability to lead, increasing more rapidly in periods of growing financial stress.  This is partly responding to changes in relationships across subsets of series that are substantial enough to impact the IS/VB quality and signal caution,  and that $K_t$ is a daily measure while the St.~Louis index is weekly.   Routine use of the entropy index as a monitor on model adequacy is recommended.

\subsection{Evaluation and Highlight of the Role of Recoupling} 

\begin{figure}[htbp!]
\centering
\includegraphics[width=\textwidth]{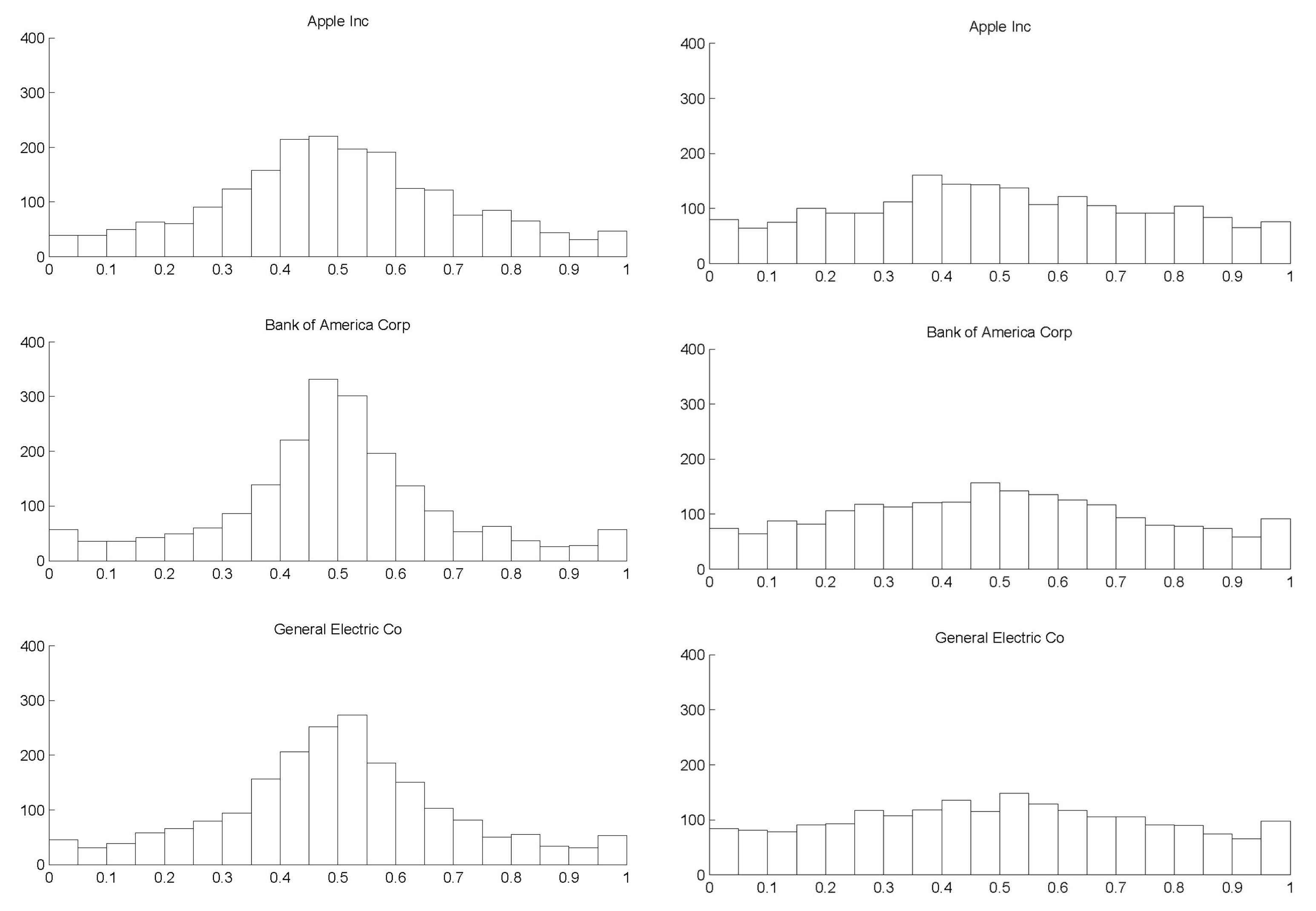}
\caption{Realized $1-$step forecast c.d.f. values for three stocks. {\em Left:} Without recoupling; {\em Right:} with recoupling. Recoupling induces a more uniform distribution consistent with model adequacy. \label{fig:3stocksrecoupeeffect}}       
\end{figure}

Questions arise as to whether the IS/VB analysis can be dropped without loss when $\bGamma_t$ is very sparse. In the S\&P analysis~\citep{GruberWest2017ECOSTA} the $401-$dimensional model is very sparse;   $|sp(j)|=20$ for each $j$ so that
95\% of entries in $\bGamma_t$ are zero.  Thus the  decoupled analysis can be expected to be close to that of a DDNM.  One assessment of whether this is tenable is based on $1-$step forecast accuracy.  In any model,  for each series $j$ and time $t$, let $u_{j,t} = P(y_{j,t}|\cD_{t-1})$  be the realized value of the $1-$step ahead forecast c.d.f. 
The more adequate the model, the closer the $u_{j,t}$ to resembling $U(0,1)$ samples; if the model generates the data, the $u_{j,t}$ will be theoretically $U(0,1).$
From the SGDLM analysis noted, Figure~\ref{fig:3stocksrecoupeeffect} shows histograms of the $u_{j,t}$ over the several years for 3 chosen series.   The figure also shows such histograms based on analysis that simply ignores the IS/VB decouple/recouple steps.   This indicates improvements in that the c.d.f. \lq\lq residuals'' are  closer to uniform with recoupling. These examples are quite typical of the 401 series; evidently,  recoupling is practically critical even in very sparse (non-triangular) models. 

\subsection{Perspectives on Model Structure Uncertainty in Prediction \label{subsec:SGDLMhotspot} } 

SGDLM analysis faces the same challenges of  parameter and parental set specification  as in special cases of DDNMs.  
Scaling presses the questions of how to assess and modify the $sp(j)$ over time, in particular.  The theoretical view that these are parameters  for an extended model uncertainty analysis leads to enormous model spaces and is simply untenable computationally.   More importantly, inference on parental set membership as parameters-- i.e., model structure \lq\lq identification''-- is rarely a goal.  As~\cite{GruberWest2016BA} and \cite{GruberWest2017ECOSTA} exemplify, a more rational view is that parental sets are choices to be made based on forecast accuracy and decision outcomes, and the interest in exploring choices should be based  on these specific goals;  often I am not at all interested in \lq\lq learning'' these aspects of model structure-- I want good choices in terms of forecast and decision outcomes.  With $q$ even moderately large,  each series $j$ 
may be adequately and equally-well predicted using one of many possible small parental sets, especially in contexts such as ours of high levels of (dynamic) interdependencies. Any one such choice is preferable to weighting and aggregating a large number  since  small differences across them simply contribute noise; hence, I focus on  \lq\lq representative'' parental sets to use as a routine, with sequential monitoring over time to continually assess adequacy and respond to changes by intervention to modify the parental sets.   

\cite{GruberWest2017ECOSTA} developed a Bayesian decision analysis-inspired approach in which $sp(j)$ has 3 subsets: a \lq\lq core set'', a \lq\lq warm-up'' set, and a ``cool-down'' set. 
A  simple Wishart discount model is run alongside the SGDLM to identify series  not currently in $sp(j)$ for potential inclusion in the warm-up set. Based on posterior summaries in the Wishart model at each time $t$, one such series is added to the warm-up subset of $sp(j)$. Also at each $t$,  one series in the current cool-down subset is moved out of $sp(j)$ and series in the warm-up subset are considered to be moved to the core subset based on current posterior assessment of predictive relationships with series $j$. Evolving the model over time allows for learning on state elements related to new parental series added, and adaptation for the existing parents removed.   This nicely enables smooth changes in structure over time via the 
warm-up and cool-down periods for potential parental predictors, avoiding the need for abrupt changes and model refitting with updated parental sets. 

\begin{figure}[ht!]
\includegraphics[width=\textwidth]{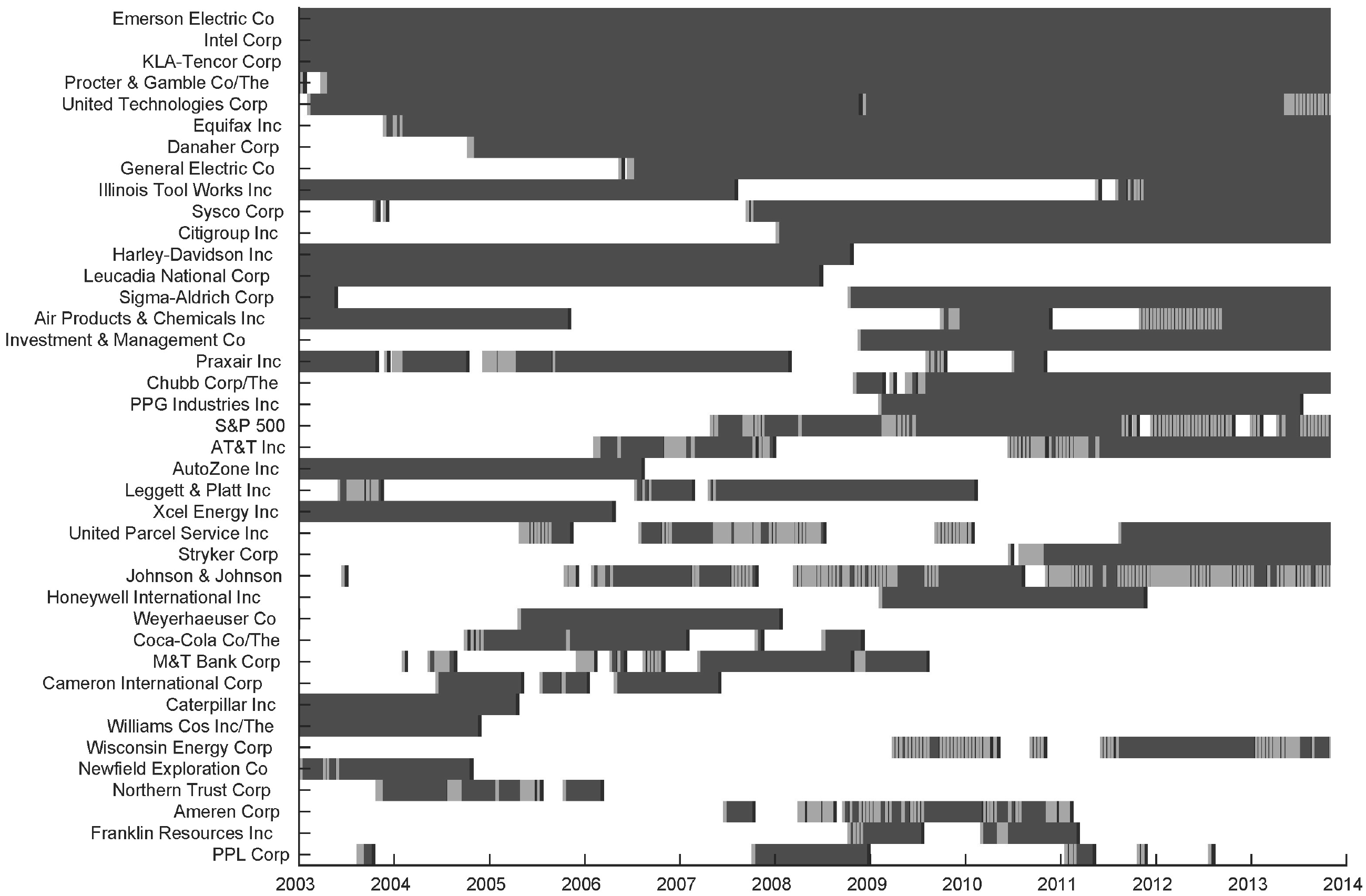}
\caption{Parental inclusion for 3M SGDLM.  {\em Dark shading}: predictor stocks included as simultaneous parents; {\em Light shading}: stocks being  considered for inclusion or to be dropped;  {\em White}: stocks not included nor under consideration. \label{fig:spinclusionimage} } 
\end{figure}

Figure~\ref{fig:spinclusionimage} gives an illustration: series $j$ is the stock price of company 3M.  Analysis fixed $|pa(j)|=20$ and allowed structural update each day; this is overkill as changes should only be made when  predicted to be beneficial, and operating over longer periods with a given model is preferable if it is deemed adequate.  That said, the figure is illuminating.  Several series are in $sp(j)$ over the entire period; several others come in/out once or twice but are clearly  relevant over time; some enter for very short periods of time, replacing others.  Then, relatively few of the 400 possible parental series are involved at all across the years. The analysis  identifies core simultaneous predictors of 3M while exploring changes related to  collinearities among others. Finally,  the names of series shown in the figure are of no primary interest. Viewing the names indicates how challenging it would be to create a serious contextual interpretation. I have little interest in that;  these are series that aid in predicting 3M price changes while contributing to quantifying multivariate structure in $\bOmega_t$, its dynamics and implications for portfolio decisions, and that is what we require per analysis goals.

\subsection{Challenges and Opportunities} 

As discussed in Section~\ref{subsec:SGDLMhotspot},  the very major challenge is that of addressing the huge model structure uncertainty problem consistent with the desiderata of (i) scalability with $q$, and (b) maintaining tractability and efficiency of the sequential filtering and forecasting analysis.  Routine model averaging is untenable computationally and, in any case,   addresses what is often a non-problem.   Outcomes in specific forecasting and/or decision analyses   should guide thinking about new ways to address this.  The specific Bayesian hot-spot technique exemplified is a step in that direction, though somewhat ad-hoc in its current implementation.  Research questions relate to broader issues of model evaluation, combination and selection and may be addressed based on related developments in other areas such as Bayesian predictive synthesis~\citep{McAlinnWest2017bpsJOE,McAlinnEtAl2017} and other methods emerging based on decision perspectives~\citep[e.g.][]{Walker2001,Clyde2012,McAlinnEtAldiscussionBA2018,Yaoetal2018}. Opportunities for theoretical research are clear, but the challenges of effective and scalable computation remain major. 

A perhaps subtle aspect of model evaluation is that, while some progress can be made at the level of each univariate series (e.g., training data to select discount factors) much assessment of forecast and decision outcomes can only be done with the recoupled multivariate model.  This should be an additional guiding concern for new approaches. 
 
SGDLMs involve flexible and adaptive models for stochastic volatility at the level of each univariate time series.  Explaining (and, in the short-term, predicting) volatility of a single series through the simultaneous parental concept is of inherent interest in itself. Then, the ability to coherently adapt the selection of parental predictors-- via the Bayesian hot-spot as reviewed in Section~\ref{subsec:SGDLMhotspot} or perhaps other methods-- opens up new opportunities for univariate model advancement. 
 
There is potential for more aggressive development of the IS/VB-based  ESS/KL measures of model adequacy with practical import.   As exemplified,  the $K_t$ entropy index relates to the entire model-- all states and volatilities across the $q$ series.   KL divergence on any subset of this large space can be easily computed, and in fact relates to opportunities to improve the IS accuracy on reduced dimensions.  This opens up the potential to explore ranges of entropy indices for subsets of series-- e.g.,  the set of industrial stocks, the set of financial/banking stocks, etc.-- separately. Changes observed in the overall $K_t$ may be reflected in states and volatilities for just some but not all stocks or sectors, impacting the overall measure and obscuring the fact that some or many components of the model may be stable. At such times, intervention to adapt models  may then be focused and restricted to only the relevant subsets of the multivariate series. 


\section{Count Time Series:   Scalable Multi-Scale Forecasting \label{sec:DBCMs} } 

\subsection{Context and Univariate Dynamic Models of Non-Negative Counts \label{sec:DBCMsbinaryPoisson}} 

Across various areas of application,  challenges arise in problems of monitoring and forecasting discrete time series, and notably many related time series of counts. These are increasingly common in areas such as consumer behavior in a range of socio-economic contexts, various natural and biological systems, and commercial and economic problems of analysis and forecasting of discrete outcomes~\citep[e.g.][]{Cargnoni1997,yelland2009bayesian,Terui2014,Chen2017,Soyer2018,Glynn2019BayesianAO}. Often there are questions of modeling simultaneously at different scales as well as of integrating information  across series and scales~\citep[chapter 16 of][]{WestHarrison1997,Ferreira2006}.  The recent, general state-space models of~\cite{BerryWest2018DCMM} and~\cite{BerryWest2018TSM} focus on such contexts under our desiderata: defining flexible, customisable models for decoupled univariate series, ensuring relevant and coherent cross-series relationships when recoupled, and  maintaining scalability and computational efficiency in sequential analysis and forecasting. The theory and methodology of such models is applicable in many fields, and define new research directions and opportunities in addressing large-scale, complex and dynamic discrete data generating systems. 

New classes of dynamic generalized linear models~(DGLMs,~\citealp[][chapter 14]{West1985a,WestHarrison1997})   include  dynamic count mixture models (DCMM, ~\citealp{BerryWest2018DCMM}) and extensions to  dynamic binary cascade models (DBCM, ~\citealp{BerryWest2018TSM}).   These exploit coupled dynamic models for binary and Poisson outcomes in structured ways. Critical advances for univariate count time series modeling include  the use of time-specific random effects to capture over-dispersion, and customized \lq\lq binary cascade'' ideas for predicting clustered count outcomes and extremes.  These developments are exemplified in forecasting customer demand and sales time series in  these papers, but are of course of much broader import.   I focus here on   the multi-scale structure and use simple conditional Poisson DGLMs as examples.   Each time series $y_{j,t} \sim Po(\mu_{j,t})$ with log link $\log(\mu_{t}) = \bF_{j,t}'\btheta_{j,t}$ where state vectors $\btheta_{j,t}$ follows linear Markov evolution models-- independently across $j$-- as in DLMs.  Decoupled, we use the traditional sequential filtering and forecasting analysis exploiting (highly accurate and efficient) coupled variational Bayes/linear Bayes 
computations~\citep{West1985a,WestHarrison1997,Triantafyllopoulos2009}.  Conditional on the $\bF_{j,t}$, analyses are decoupled across series.    

\subsection{Common Dynamic Latent Factors and Multi-Scale Decouple/Recouple}     
    
Many multivariate series share common patterns or effects for which hierarchical or traditional dynamic latent factor models would be first considerations. Integrating hierarchical structure into dynamic modeling has seen some development~\citep[e.g.][]{GamermanMigon1993,Cargnoni1997,FerreiraGamermanMigon1997}, but application quickly requires intense computation such as MCMC and obviates efficient sequential analysis and scaling to higher dimensions  with more structure across series.  The same issues arise with dynamic latent factor models, Gaussian or otherwise~\citep[e.g.][]{LopesCarvalho07,Carvalho11,NakajimaWest2013JFE,kastner2017,NakajimaWest2017BJPS,McAlinnEtAl2017}.  The new multi-scale approach of~\cite{BerryWest2018DCMM} resolves this with novel Bayesian model structures that define latent factor models but maintain fast sequential analysis and scalability.  The ideas are general and apply to all dynamic models,  but are highlighted here in the conditional Poisson DGLMs.   Suppose that series $j$ has
$\bF_{j,t}' = (\bx_{j,t}',\bphi_t') $ where $\bx_{j,t}$ include series $j$-specific predictors and $\bphi_t$ represents a vector of dynamic latent factors impacting all series.  The state vectors are conformably partitioned: $ \btheta_{j,t}'=(\bgamma_{j,t}',\bbeta_{j,t}')$  where $\bbeta_{j,t}$ allows for diversity of the impact of the latent factors across series.  

Denote by  $\cM_j$ the DGLM for series $j.$  With independent priors on states across series and conditional on latent factors $\bphi_{\seq t{t+h}}$ over $h-$steps ahead,  analyses are decoupled: forward filtering and forecasting for the $\cM_j$ are parallel and efficient.  The multi-scale concept involves an external or \lq\lq higher level/aggregate'' model $\cM_0$  to infer and predict the latent factor process, based on \lq\lq top-down'' philosophy~\citep[][section 16.3]{WestHarrison1997}. That is, $\cM_0$ defines a current posterior predictive distribution for $\bphi_{\seq t{t+h}}$ that feeds each of the $\cM_j$ with values for their individual forecasting and updating.   Technically, this uses forward simulation:  $\cM_0$  generates Monte Carlo samples of  latent factors, and for every such sample, each of the decoupled $\cM_j$ directly updates and forecasts.  In this way,  informed predictions of latent factor processes from $\cM_0$ lead to fully probabilistic inferences at the micro/decoupled series level, and within each there is an explicit accounting for uncertainties about the common features $\bphi_{\seq t{t+h}}$ in the resulting series-specific analyses.

\subsection{Application Contexts, Model Comparison and Forecast Evaluation} 

\begin{figure}[t!] 
 \centering
\includegraphics[width=.80\textwidth]{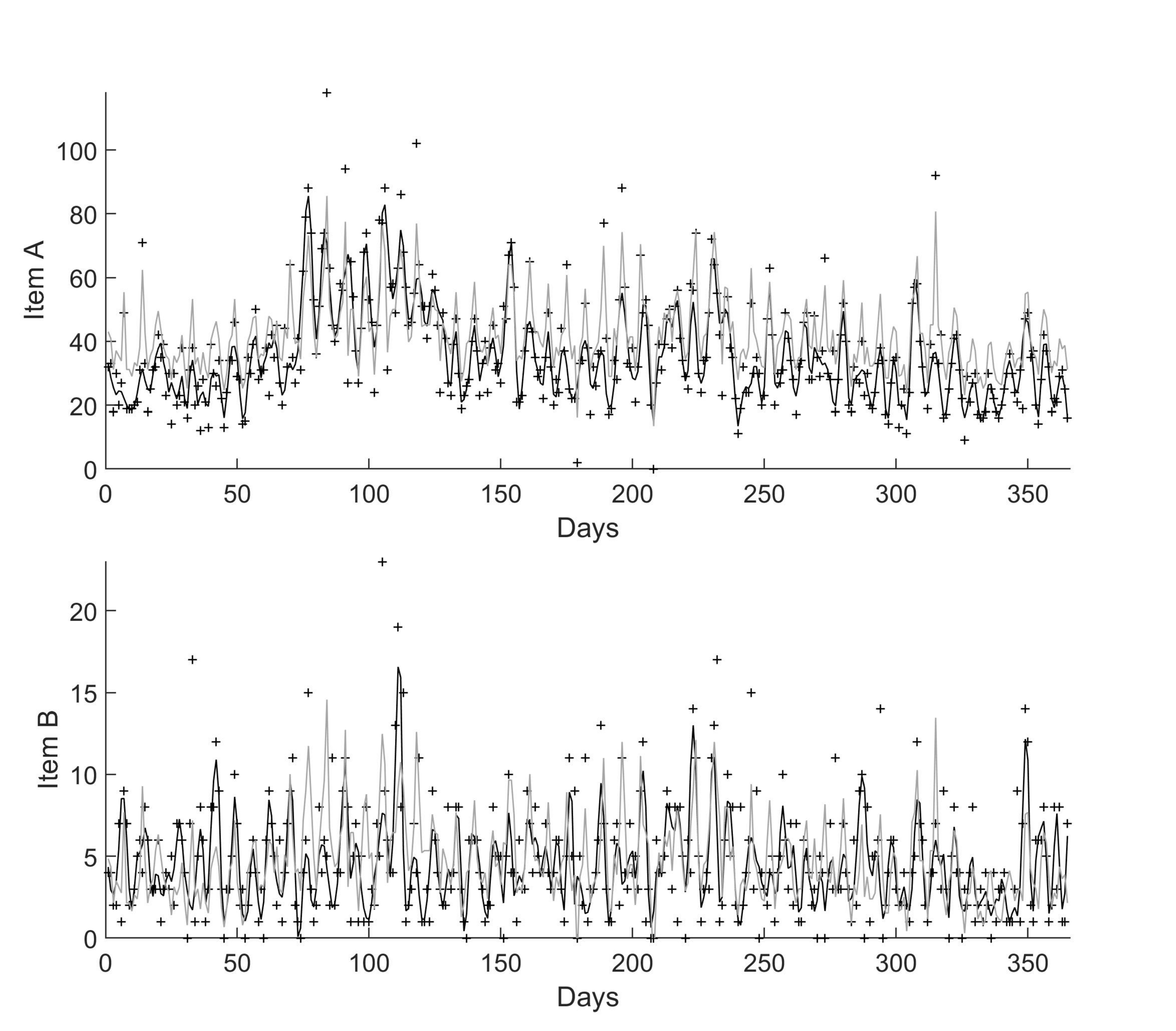}
\caption{Sales data on two pasta items in one store over 365 days, taken from a large case study in~\cite{BerryWest2018TSM}.   Daily data are {\small +};  black lines indicate item-specific day-of-week seasonal structure, while the grey line is that from an aggregate model $\cM_0.$
Item-specific effects appear as stochastic variations on the latter, underscoring interest in information sharing via a multi-scale analysis.  Diverse levels and patterns of stochastic variation apparent are typical across many items; item A is at high levels, item B lower with multiple zeros. This requires customized components in each  of the decoupled univariate dynamic models, while improved forecasts are achieved via multi-scale recoupling. 
\label{fig:pastasalesdata2eg}   }
\end{figure}

Supermarket sales forecasting examples~\citep{BerryWest2018DCMM,BerryWest2018TSM} involve thousands of individual items across many  stores,  emphasizing needs for efficiency and scalability of analyses.  The focus is on  daily transactions and sales data in each store:  for each item, and over multiple days ahead to inform diverse end-user decisions in supply chain management and at the store management level. Models involve item-level price and promotion predictors, as well as critical day-of-week seasonal effects.  The new univariate models allow for diverse levels of sales,  over-dispersion via dynamic random effects, sporadic sales patterns of items via dynamic zero-inflation components, and rare sales events at higher levels.   Daily seasonal patterns are a main focus for the new multi-scale approach.   In any store, the \lq\lq traffic'' of for, example, the overall number of customers buying some kind of pasta product is a key predictor of sales of any specific pasta item;     hence an aggregate-level $\cM_0$ of total  sales-- across all pasta items-- is expected to define more accurate evaluation and prediction of the seasonal effects for any one specific item  than would be achievable using only day on that item.  Figure~\ref{fig:pastasalesdata2eg} displays two example sales series; these illustrate  commonalities as well as noisy, series-specific day-of-week structure and other effects (e.g., of prices and promotions).  
Given very noisy data per series but inherently common day-of-week traffic patterns, this is an ideal  context for the top-down, multi-scale decouple/recouple strategy. 

Results in~\cite{BerryWest2018TSM} demonstrate advances in  statistical model assessments and in terms of measures of practical relevance in the consumer demand and sales context. 
A key point here is that the very extensive evaluations reported target both statistical and contextual concerns: (a) broad statistical evaluations include assessments of frequency calibration (for binary and discrete count outcomes) and coverage (of Bayesian predictive distributions), and their comparisons across models;  (b)  broad contextual evaluations  explore ranges of metrics to evaluate specific models and compare across models--  metrics based on  loss functions such as mean absolute deviation,  mean absolute percentage error, and others  that are industry/application-specific and  bear on practical end-user decisions.  These studies represent a focused context for advancing the main theme that model evaluation should be arbitrated in the contexts of specific and explicit forecast and decision goals in the use of the models. Purely statistical evaluations are required as sanity checks on statistical model adequacy, but only as precursors to the defining concerns in applying models.

\subsection{Challenges and Opportunities} 

The DCMM and DBCM frameworks define opportunities for applications in numerous areas-- such of monitoring and forecasting in  marketing and consumer behavior contexts, epidemiological studies, and others where counts arise from underlying complex, compound and time-varying processes, In future applications, the shared latent factor processes will be multivariate, with dimensions reflecting different ways in which series are conceptually related. The new multi-scale modeling concept and its decouple/recouple analysis opens up potential to apply to many areas in which there are tangible aggregate-level or other, external information sources that generate information relative to aspects of the common patterns/shared structure in multiple series.  One the challenges is that, in a given applied context, there may be multiple such aggregate/higher-level abstractions, so that technical model developments will be of interest to extend the analysis to integrate inferences (in terms of \lq\lq top down'' projections'') from two or more external models.    A further challenge and opportunity relates to the question of maintaining faith with the desiderata of fast and scalable computation; the approaches to date involve extensive-- though direct-- simulation in $\cM_0$ of the latent factors $\bphi_t$ for projection to the micro-level models $\cM_j.$ In extensions with multiple higher-level models, and with increasing numbers $q$ of the univariate series within each of which concomitant simulations will be needed, this will become a computational challenge and limitation.  New theory and methodology to address these coupled issues in scalability are of interest.

\section{Multivariate Count Series:   Network Flow Monitoring \label{sec:DYNETs} } 
 
\subsection{Dynamic Network Context and DGLMs for Flows} 

\begin{figure}[t!]
$$
\xymatrix{ 
\textrm{Entering network} \ar@/_0pc/@{>}^*+{y_{0,i,t}}[rrrd]  & & & & & & *+[F-]{1} \ar@{.}[d] & \\
& & &
	*+[F-]{i} \ar@(ul,ur)^*{y_{i,i,t}}     \ar@/^/@{>}^*+{y_{i,1,t}}[rrru] 
	\ar@{>}^*+{y_{i,j,t}}[rrr] \ar@/_/@{>}^*+{y_{i,I,t}}[rrrd] 
	\ar@/_0pc/@{>}_*+{y_{i,0,t}}[llld] & & & *+[F-]{j} \ar@{.}[d] & \\
 \textrm{Leaving network} & & & & & & *+[F-]{I} & \\
}
$$ 
\caption{Network schematic and notation for flows at time $t$.   \label{fig:networkcartoon} }
\end{figure}
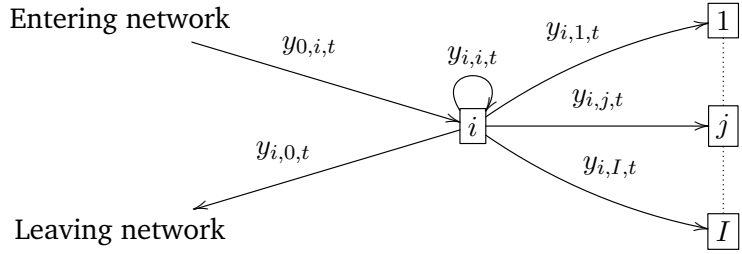

Related areas of  of large-scale count time series concern flows of \lq\lq traffic'' in various kinds of networks. 
This topic is significantly  expanding with increasingly large-scale data in  internet and social network contexts, and with regard to 
physical network flow problems.   Bayesian   models have been developed for  network tomography and physical
traffic flow forecasting~\citep[e.g.][]{Tebaldi1998,Cogdon2000,Tebaldi2002,Queen13,JandarovEtAl2014,Hazelton2015}, but increasingly large dynamic network flow problems require new modeling approaches.  I contact  recent innovations that address: (a) scaling of flexible and adaptive models for analysis of large networks to characterize the inherent variability and stochastic structure in flows between nodes, and into/out of networks; (b)  evaluation of formal statistical metrics to monitor dynamic network flows and signal/allow for 
informed interventions to adapt models in times of signalled change or anomalies; and (c) evaluation of inferences on subtle aspects of dynamics in network structure related to node-specific and node-node interactions over time that also scale with network dimension.   
These goals interact with the core desiderata detailed earlier of statistical and computational efficiency, and scalability of Bayesian analysis, with the extension of doubly-indexed count time series:  now, $y_{i,j,t}$ labels the count of traffic (cars, commuters, IP addresses, or other units) \lq\lq flowing'' from a node $i$ to a node $j$ in a defined network on $I$ nodes in time interval $t-1\to t$; node index  0 represents \lq\lq outside'' the network as in Figure~\ref{fig:networkcartoon}.

In dynamic network studies of various kinds, forecasting may be of interest but is often not the primary objective. More typically,  the goals are to characterize normal patterns of stochastic variation in flows, monitor and adapt models to respond to changes over time, and inform decisions based on signals about patterns of changes.   Networks are increasingly large;  internet and social networks can involve hundreds or thousands of nodes, and are effectively unbounded in any practical sense from the viewpoint of statistical modeling. The conceptual and technical innovations in~\cite{Chenetal2018JASA} and~\cite{ChenBanksWest2018}  define flexible multivariate models exploiting two developments of the  decouple/recouple concept-- these  advance the ability to address the above concerns in a scalable Bayesian framework.

 
\subsection{Decouple/Recouple for Dynamic Network Flows} 
Dynamic models in~\cite{Chenetal2018JASA} and~\cite{ChenBanksWest2018} use flexible, efficient Poisson DGLMs for in-flows to the network $y_{0,i,t}$ independently across nodes $i=\seq 1I.$ Within-network flows are inherently conditionally multinomial  i.e., $y_{i,\seq 0I,t}$ is multinomial based on the current \lq\lq occupancy'' $n_{i,t-1}$ of node $i$ at time $t$ . The first use of decoupling is to break the multinomial into a set of $I$ Poissons, taking 
$y_{i,j,t} \sim Po(m_{i,t}\phi_{i,j,t}) $ where $\log(\phi_{i,j,t}) = \bF_{i,j,t} \btheta_{i,j,t}$ defines a Poisson DGLM  with state vector $\btheta_{i,j,t}.$
The term $m_{i,t} = n_{i,t-1}/n_{i,t-2}$ is an offset to adjust for varying occupancy levels. With  independence across nodes,  this yields a set of $I+1$ Poisson DGLMs per node that are decoupled for on-line learning about underlying state vectors.   Thus fast, parallel analysis yields posterior inferences on the $\phi_{i,j,t}$; Figure~\ref{fig:dynetexample}(a) comes from an example discussed further in Section~\ref{subsec:Dynetsexample}.   Via decoupled posterior simulation, these are trivially mapped to implied transition probabilities in the node- and time-specific multinomials implied, i.e., for each node $i$, the probabilities  $\phi_{i,j,t}/\sum_{j=\seq 0I}\phi_{i,j,t}$ on $j=\seq 0I.$  

\subsection{Recoupling for Bayesian Model Emulation} 
The second use of recoupling   defines an approach Bayesian to “model emulation”~\citep[e.g.][]{Liu2009,IrieWest2018portfoliosBA} in the dynamic context.  While the decoupled DGLMs run independently,  they are able to map relationships across sets of nodes as they change over time.  Using posterior samples of trajectories of the full sets of $\phi_{i,j,t}$,  we are   able to emulate inferences in a more structured model that explicitly involves node-node dependencies. 
Specifically, the so-called dynamic gravity models (DGM) of~\cite{Chenetal2018JASA} and~\cite{ChenBanksWest2018} extend prior ideas of two-way modeling in networks and other areas~\citep[e.g.][]{West1994, Sen:1995,Cogdon2000} to a rich class of dynamic interaction structures. The set of modified Poisson rates are mapped to a DGM via  $\phi_{i,j,t} = \mu_t\alpha_{i,t} \beta_{j,t} \gamma_{i,j,t}$ 
where: (i) $\mu_t$  is an overall network flow intensity process over time, (ii)  $\alpha_{i,t}$ is a node $i$-specific \lq\lq origin (outflow)''  process, (iii) $\beta_{j,t}$ is a node $j$-specific  \lq\lq destination (inflow)''  process, and (iv) $\gamma_{i,j,t}$ is a node $i\to j$ 
 \lq\lq affinity  (interaction)''  process.  Subject to trivial aliasing constraints (fixing geometric means of main and interaction effects at 1) this is an invertible map between the flexible decoupled system of models and the DGM effect processes.    

\subsection{Application Context and On-line Model Monitoring for Intervention \label{subsec:Dynetsexample} } 

\begin{figure}[b!]
\centering
\includegraphics[width=\textwidth]{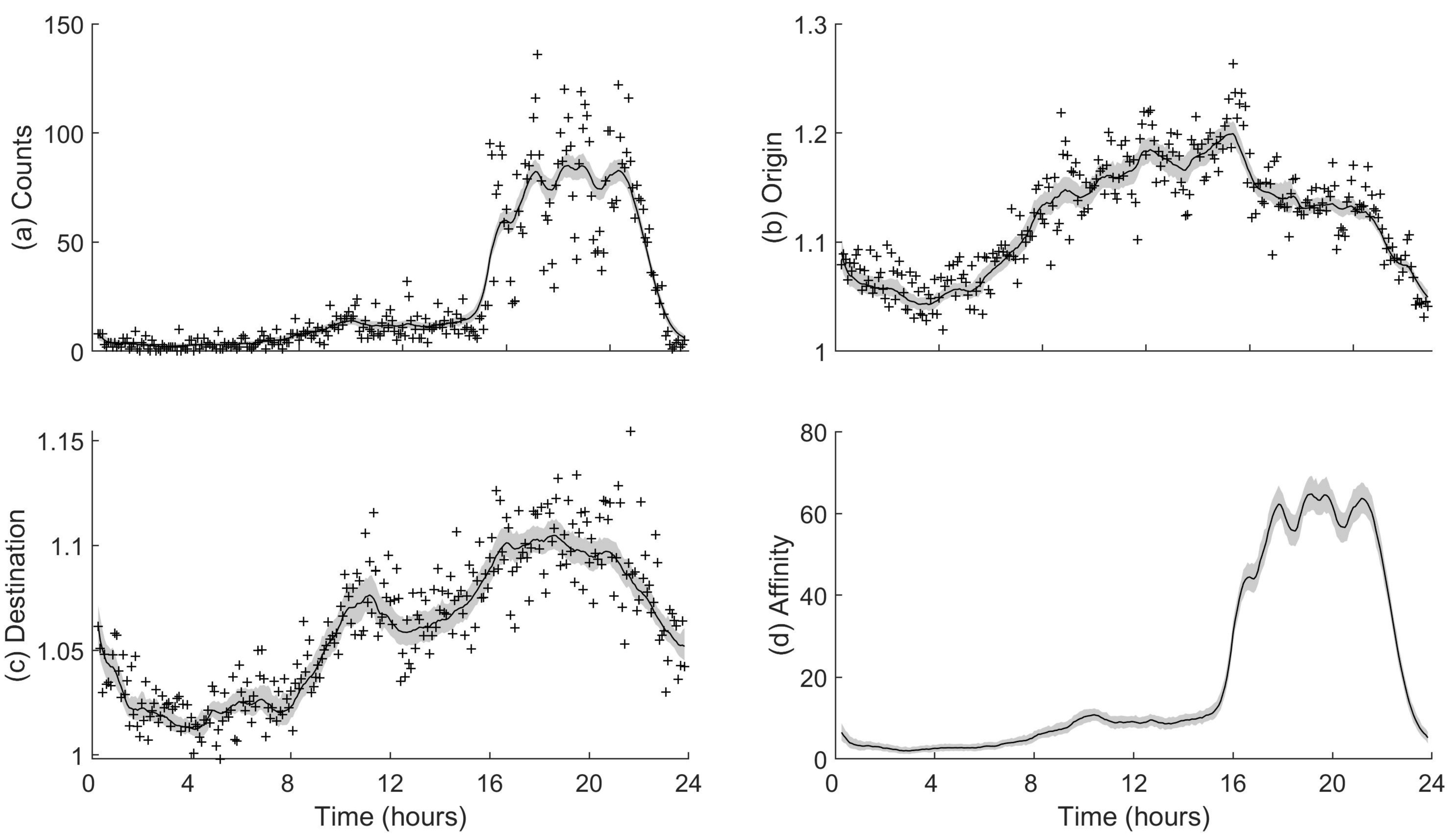}
\caption{Posterior summaries for aspects of  flows involving two web domain nodes in the Fox News web site on September 17, 2015. 
Nodes $i=$ Games/Online Games and $j=$ Games/Computer \& Video Games. 
(a) Posterior trajectory for Poisson levels $\phi_{i,j,t}$ for flows $i\to j$; 
(b) Posterior trajectory for the origin (outflow)  process $\alpha_{i,t}$;
(c) Posterior trajectory for the destination (inflow)  process $\beta_{j,t}$; 
(d) Posterior trajectory for the affinity process $\gamma_{i,j,t}$.   Trajectories are approximate posterior means and $95\%$ credible interval, and the 
{\small +} symbols indicate empirical values from the raw data.
 \label{fig:dynetexample}   }    
\end{figure}

Case studies in~\cite{ChenBanksWest2018} concerns flow data recording individual visitors (IP addresses) to well-defined nodes (web ``domains'') of the Fox News website.  Studies include a   network of  $I=237$ nodes illustrating scalability (over $56{,}000$ node-node series). 
Counts are for five-minute intervals, and key examples use data on September 17th, 2015; see Figure~\ref{fig:dynetexample} looking at flows from 
node $i=$\lq\lq Games/Online Games'' and $j=$\lq\lq Games/Computer \& Video Games'', with raw flow counts in frame (a). 

In this example, there are no relevant additional covariates available, so the univariate Poisson DGLMs are taken as local linear trend models, with  $2-$dimensional state vectors representing local level and gradient at each time~\citep[][chapt. 7]{WestHarrison1997}.  While this is a flexible model for adapting to changes in the $\phi_{i,j,t}$ over time as governed by model discount factors,  it is critical to continuously monitor model adequacy over time in view of the potential for periods when flows represent departure from the model, e.g., sudden unpredicted bursts of traffic or unusual decreases of traffic over a short period based on news or other external events not available to the model. This aspect of model evaluation is routine in many other areas of Bayesian time series and there are a range of technical approaches. Arguably most effective-- and certainly analytically and computationally almost trivial-- is  Bayesian model monitoring and adaptation methodology based on sequential Bayes' factors as tracking signals in a decision analysis context~(\citealp{West1986,WestHarrison1997} chapt. 11). 
Each DGLM   is subject to such automatic monitoring and the ability to adapt the model via flagging outliers and using temporarily decreased discount factors to more radically adapt to structural changes. In  Figure~\ref{fig:dynetexample}, it can be seen that this is key in terms of two   periods of   abrupt changes in gradient of the local linear trend around the 16 and 22 hour marks, and then  again for a few short periods later in the day when flows are at high levels but exhibit swings up/down. 

Figure~\ref{fig:dynetexample} also shows trajectories of imputed DGM processes from recoupling-based emulation.  Here it becomes clear that both the node $i$ origin and node $j$ destination effect processes vary through the day, with the latter increasing modestly through the afternoon and evening, and then they each decay at later hours.  Since these processes are multipliers in the Poisson means and centered at 1, both origin and destination processes   represent flow effects above the norm across the network.  The figure also shows the trajectory of the affinity effect process $\gamma_{i,j,t}$ for these two nodes.  Now it becomes quite clear that the very major temporal pattern is idiosyncratic to these two nodes;   the interaction process boosts very substantially at around the 16 hour mark, reflecting domain-specific visitors at the online games node aggressively flowing to the computer and video games node in the evening hours.

 \subsection{Challenges and Opportunities} 

The summary example above and  more in~\cite{ChenBanksWest2018} highlight the utility of the new models and the decouple/recouple strategies. 
Critically, DGMs themselves are simply not amenable to fast and scalable analysis; the recouple/emulation method enables scalability (at the optimal  
rate ${\sim}I^2$) of inferences on what may be very complex patterns of interactions in flows among nodes as well as in their origin and destination main effects.   
For future applications, the model is open to use of node-specific and node-node pair covariates in underlying univariate DGLMs when such information is available.   Analysis is also open to the use of feed-forward intervention information~(\citealp{West1989,WestHarrison1997} chapt. 11) that may be available to anticipate upcoming changes that would otherwise have to be signalled by automatic monitoring.  Canonical Poisson DGLMs can be extended to richer and more flexible forms; without loss in terms of maintaining faith with the key desiderata of analytic tractability and computational efficiency, the  models in Section~\ref{sec:DBCMsbinaryPoisson}
offer potential to improve characterisation of patterns in network flows via inclusion of dynamic includes random effects for over-dispersion as well as flexible models for very low or sporadic flows between certain node pairs.   
Finally,  these models and emulation methods will be of interest in applications in areas beyond network flow studies. 
     


\newpage

 \def\zero{\mbox{\boldmath$0$}}
 \def\b{\mbox{\boldmath$b$}}
 \def\x{\mbox{\boldmath$x$}}
 \def\y{\mbox{\boldmath$y$}}
 \def\A{\mbox{\boldmath$A$}}
 \def\B{\mbox{\boldmath$B$}}
 \def\F{\mbox{\boldmath$F$}}
 \def\bepsilon{\mbox{\boldmath$\varepsilon$}}

\newcommand{\omegabf}{\boldsymbol{\omega}}
\newcommand{\pibf}{\boldsymbol{\pi}}
\newcommand{\phibf}{\boldsymbol{\tilde{\phi}}}
 \def\one{\mbox{\boldmath$1$}}
 \def\bGamma{\mbox{\boldmath$\Gamma$}}
 

 \setcounter{page}{1}\thispagestyle{empty}
\begin{center} 
{\LARGE\bf\blu  Discussion of \\ ``Bayesian Forecasting of Multivariate Time Series: 
\smallskip

Scalability, Structure Uncertainty and Decisions''}
  
\bigskip\bigskip
{\Large Chris Glynn\\} Assistant Professor of Decision Sciences\\ Peter T. Paul College of Business and Economics\\ University of New Hampshire, U.S.A.\\ 

\bigskip
\href{mailto:christopher.glynn@unh.edu}{christopher.glynn@unh.edu}

\bigskip
\today
\end{center} 
 
I congratulate Professor West for his 2018 Akaike Memorial Lecture Award and for articulately synthesizing recent research in this unified treatment of the ``decouple/recouple'' framework.  For readers that learned Bayesian dynamic models from \cite{westharrison}, the motivation and multivariate extensions of univariate dynamic linear models (DLMs) are familiar.  The current focus on modeling sparse cross-series structure for scaling posterior computations to high dimensions is a welcome addendum.

A current challenge in Bayesian analysis is to scale models and computational procedures to meet the demands of increasingly large and complex data without sacrificing fundamentals of applied statistics. Physical, social, and economic sciences rely heavily on statistical models that are richly structured, interpretable, and reliably quantify uncertainty.  There is great value to science in models that are both interpretable \textit{and} scalable.  In this regard, the decouple/recouple framework is an important contribution for modeling large collections of time series.

The computational gains of the decouple/recouple framework are achieved by exploiting sparse cross-series structure. At each time $t$, it is assumed that series $j$ has a known set of simultaneous parents, denoted $sp(j) \subseteq \{1:q\} \setminus \{j\}$. In the SGDLM setting, the observation of series $j$ at time $t$ is modeled as the composition of regressions on series-specific covariates $\x_{j,t}$ and simultaneous parental series $\y_{sp(j),t}$,
\begin{align}
    y_{j,t} = \x'_{j,t}\phi_{j,t} + \y'_{sp(j),t}\gamma_{j,t} + \nu_{j,t}.
\end{align}
The $\gamma_{j,t}$ state vector is augmented by zeros to form the $j^{th}$ row in matrix $\bGamma_t$, which encodes the joint collection of simultaneous parent relationships across all series at time $t$.  In $\bGamma_t$, the $jk^{th}$ element is non-zero if $y_{k,t}$ is a parent of $y_{j,t}$, and the analysis assumes independent Gaussian prior distributions for each non-zero coefficient.  The prior distribution is then 
\begin{align}
    \gamma_{j,k,t} \sim N(m_\gamma, \sigma^2_{\gamma}) \one_{\{k \in sp(j)\}} + \left(1-\one_{\{k \in sp(j)\}}\right) \delta_0(\gamma_{j,k,t}), \mbox{ when } j \neq k.
\end{align}  

\begin{figure}[h]
    \centering
    \begin{subfigure}{.35\textwidth}
        \centering
        \includegraphics[width = 1\textwidth]{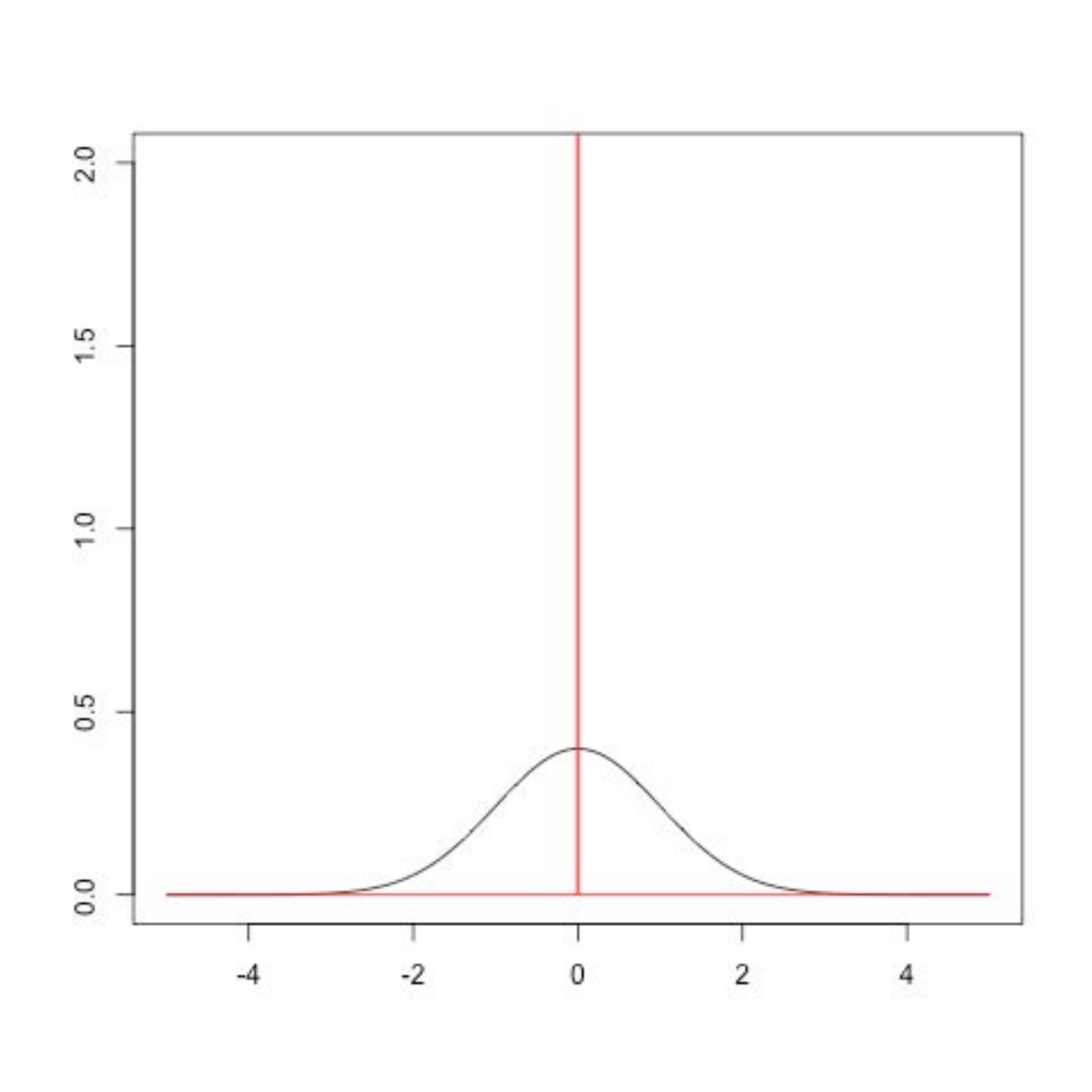}
        \caption{Point mass at zero}
        \label{subfig:point_mass}
    \end{subfigure}
    \begin{subfigure}{.35\textwidth}
        \centering
        \includegraphics[width = 1\textwidth]{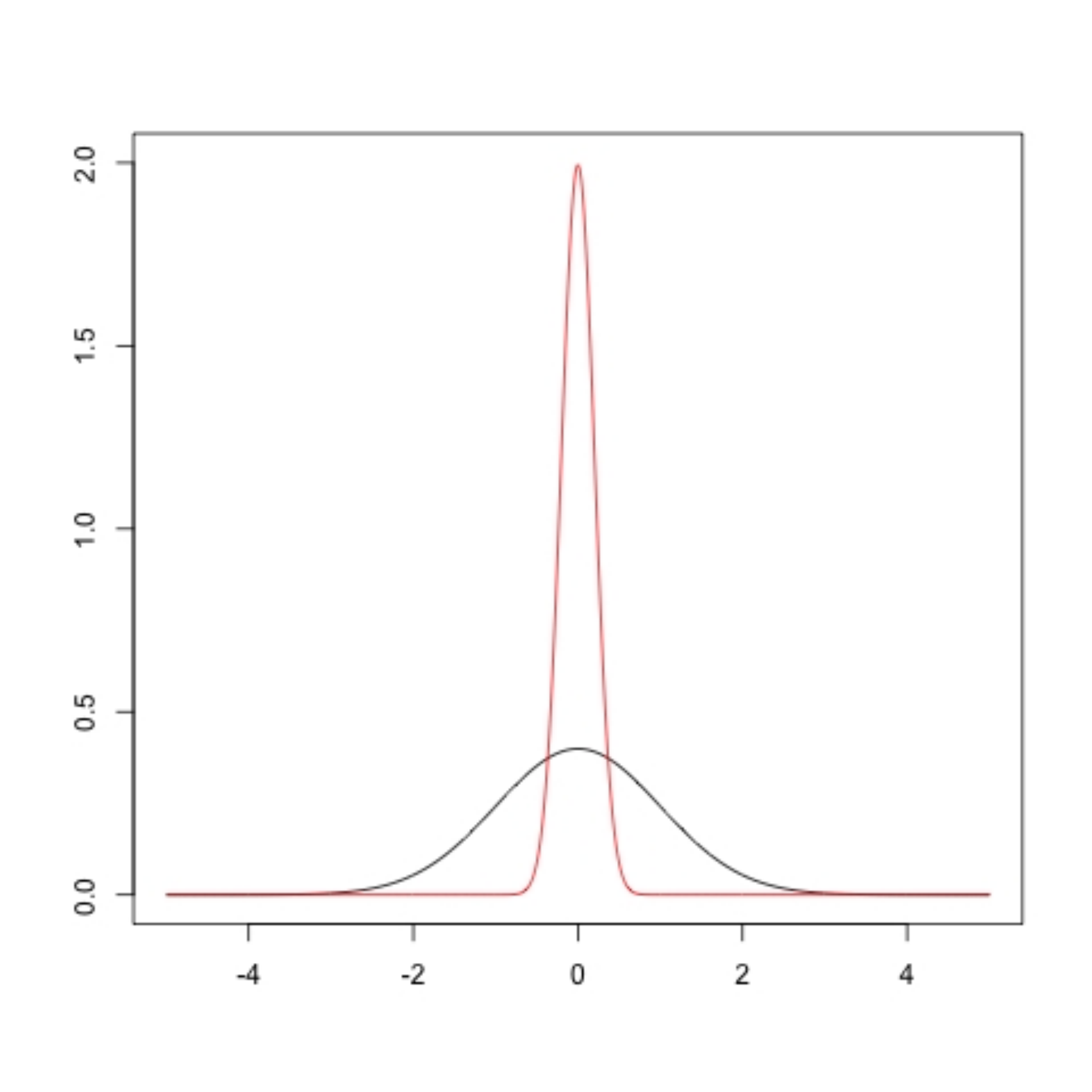}
        \caption{Spike and slab}
        \label{subfig:Spike_Slab}
    \end{subfigure}
    \caption{Mixture prior distribution with a N(0,1) component a point mass at zero (left) and spike and slab prior (right).}
    \label{fig:Mixture}
\end{figure}

While practical computational considerations in the SGDLM framework require that the set of simultaneous parents is either known or estimated with a heuristic algorithm \citep{GruberWest2017ECOSTA}), modeling the probability that series $k$ is included in $sp(j)$ sheds light on connections between the decouple/recouple framework and other well-known variable selection methods.  In addition, it points to interesting directions of future research.  Suppose a model extension where $P(k \in sp(j)) = \pi$ is the prior probability that series $k$ is in $sp(j)$.  Then the prior distribution for state variable $\gamma_{j,k,t}$ would be a mixture
\begin{align}
    \gamma_{j,k,t} \sim \pi N(m_\gamma, \sigma^2_{\gamma}) + (1-\pi)\delta_{0}(\gamma_{j,k,t}) 
\end{align}
where one component is the standard $N(m_{\gamma},\sigma^2_{\gamma})$ prior when $k \in sp(j)$ and the other is a point mass at zero when $k \notin sp(j)$.  This scenario is illustrated in Figure \ref{subfig:point_mass}.  

An interesting direction of future research is to relax the assumption that $\gamma_{j,k,t}=0$ when $k \notin sp(j)$ and introduce a Gaussian noise centered at zero instead. The exact zero that encodes sparse structure is elegant; however, there may be further computational gains to be achieved by allowing contributions from series that are \textit{approximately} rather than exactly zero.  In this spike and slab type setting \citep{George1993}, the prior for each $\gamma_{j,k,t}$ is a mixture of two Gaussian distributions, the original Gaussian component and a Gaussian component tightly concentrated around zero (Figure \ref{subfig:Spike_Slab}).  When utilizing simultaneous values from other time series $\y_{sp(j),t}$ as regressors, choosing which series to include in $sp(j)$ is a dynamic variable selection problem, and  \cite{rockova2017dynamic} utilize the spike and slab prior to model dynamic sparse structure.  

While Professor West makes it clear that, in the present work, the use of graphical structure is a means to improve forecast performance in multivariate time series and not a key inference goal, there are applications where inferring parental relationships in cross-series structure is important. One example is managing the risk of contagion in financial crises.  Given a large collection of real time stock price data for systemically important financial institutions,  inferring simultaneous parents of individual institutions (Wells Fargo, for example) is useful to both regulators and policymakers.  Learning simultaneous parents is especially important when designing market interventions to prevent (or halt) contagion.  Rather than making investments in all systemically important banks, as the Federal Reserve did in the financial crisis of 2008, a central bank could make targeted investments in the few firms that are simultaneous parents to many other institutions. 


\newpage\setcounter{page}{1}\thispagestyle{empty}
\begin{center} 
{\LARGE\bf\blu  Discussion of \\ ``Bayesian Forecasting of Multivariate Time Series: 
\smallskip

Scalability, Structure Uncertainty and Decisions''}
  
\bigskip\bigskip
{\Large  Jouchi Nakajima}\\  Senior Economist\\ Bank of Japan, Tokyo, Japan\\

\bigskip
\href{mailto:jouchi.nakajima@gmail.com}{jouchi.nakajima@gmail.com}

\bigskip
\today
\end{center} 
 
\begin{abstract}
The author focuses on the ``decoupling and recoupling'' idea  that can critically increase both computational and forecasting efficiencies in practical problems for economic and financial data. My discussion is two-fold. First, I briefly describe the idea with an example of time-varying vector autoregressions (TV-VAR), which are widely used in the context. Second, I highlight the issue of how to assess patterns of simultaneous relationships.

\medskip
\noindent {\em Keywords:} Bayesian forecasting, decouple/recouple, time-varying vector autoregressions, multivariate time series models
\end{abstract}


\section{Introduction}

I thank the author for a great discussion of recent advances in Bayesian multivariate time-series modeling strategies with several relevant and practical examples in economics and financial data problems. I believe that his comprehensive description of key model structure and methods as well as notes on challenges and opportunities are all beneficial to readers. One of the main focuses in the paper is the decoupling and recoupling idea for estimating and forecasting multivariate time-series models. For high-dimensional problems, in particular, the idea is one of the strengths of the Bayesian approach. To review it, I briefly describe an example of time-varying vector autoregressions (TV-VAR) and see how the idea is applied to the model in a practical setting. Then, I discuss the issue of simultaneous relationships that is one of the important aspects in the decoupling and recoupling strategy.


\section{An example: Time-varying vector autoregressions}

The VAR models have been popular workhorses in macro- and financial-econometrics, and the time-varying versions, TV-VAR models, have become quite popular since \cite{Primiceri05} developed a seminal form of the TV-VAR with stochastic volatility. Yet, the model structure itself was not new: it simply forms a traditional dynamic linear model~\citep[e.g.,][]{WestHarrison97}. The Primiceri's model, specialized for an analysis with macroeconomic variables, fits a variety of contexts well, in particular, fiscal and monetary policy discussions~\citep[see also][]{Nakajima11}. In financial econometrics, \cite{DieboldYilmaz09} exploit the VAR model to assess spillover effects among financial variables such as stock price and exchange rates, and \cite{GeraciGnabo18} extend the framework with the TV-VAR.

Define a response $\y_t$, ($t=1,2,\ldots$), as the $q\times 1$ vector. The TV-VAR$(p)$ model forms
        \begin{eqnarray*}
        \A_t\y_t \,=\, \sum_{j=1}^p \F_{jt}\y_{t-j} + \bepsilon_t,\quad
        \bepsilon_t \sim N(\zero, \bLambda_t),
        \end{eqnarray*}
where $\F_{jt}$ is the $q\times q$ matrix of lag coefficients, and $\bLambda_t$ is the $q\times q$ diagonal volatility matrix with $i$-th diagonal element denoted by $\sigma_{it}^2$. Note that the model can include time-varying intercepts and regression components with other explanatory variables, although these additional ingredients do not change the following discussion.

The $\A_t$ is the $q\times q$ matrix that defines simultaneous relationship among $q$ variables, which is analogous to simultaneous parents and \textit{parental predictors} in the author's discussion. With the diagonal structure of $\bLambda_t$, the $\A_t$ defines patterns of contemporaneous dependencies among the responses $\{y_{1t},\ldots,y_{qt}\}$. For identification, the model requires at least $q(q-1)/2$ elements in the off-diagonal part of $\A_t$ set to be zero.

A typical assumption for the contemporaneous structure in macroeconomic- and financial-variable data contexts is a triangular matrix:
        \begin{eqnarray*}
        \A_t \,=\, \left( \begin{array}{cccc}
        1 & 0 & \cdots & 0 \\
        -a_{21t} & \ddots & \ddots & \vdots \\
        \vdots & \ddots & \ddots & 0 \\
        -a_{q1t} & \cdots & -a_{q,q-1,t} & 1 \\
        \end{array} \right).
        \end{eqnarray*}
This leads to an implied reduced model form:
        \begin{eqnarray}
        \y_t \,=\, \sum_{j=1}^p \B_{jt}\y_{t-j} + \bnu_t,\quad
        \bnu_t \sim N(\zero, \bSigma_t),  \label{eq:model}
        \end{eqnarray}
where $\B_{jt} = \A_t^{-1}\F_{jt}$, for $j=1:p$, and $\bSigma_t = \A_t^{-1}\bLambda_t{\A_t'}^{-1}$. We can see that the variance matrix of the innovation, $\bSigma_t$, forms a Cholesky-style decomposition with $\A_t$ and $\bLambda_t$. This restricts $q(q-1)/2$ elements in $\A_t$ to be zero, and so requires no additional constraints for identification. The parental predictors of DDNMs (in Section 3) have the same structure as the contemporaneous relationship relies on only one side (upper or lower) of the triangular part in $\A_t$. The discussion of the DDNMs assumes more sparse structure as $q$ increases, i.e., most of $a_{ijt}$'s are potentially zero.

A decoupling step is implemented by recasting the model as a triangular set of univariate dynamic regressions:
        \begin{eqnarray*}
        y_{1t} &=& \b_{1t}'\x_{t-1}  + \varepsilon_{1t},  \\
        y_{2t} &=& a_{21t}y_{1t} + \b_{2t}'\x_{t-1}  + \varepsilon_{2t}, \\
        y_{3t} &=& a_{31t}y_{1t} + a_{32t}y_{2t}  +  \b_{3t}'\x_{t-1}  + \varepsilon_{3t},  \\
        \vdots\  & & \quad \vdots \\
        y_{qt} &=& a_{q1t}y_{1t} + \cdots + a_{q,q-1,t}y_{q-1,t} + \b_{qt}'\x_{t-1}\varepsilon_{qt},
        \end{eqnarray*}
where $\x_{t-1}$ is the $pq\times 1$ vector of lagged responses, defined by $\x_{t-1}'= (\y_{t-1}',\ldots,$ $\y_{t-p}')$; $\b_{it}$ is the corresponding vector that consists of lag coefficient elements in $\B_{jt}$'s; and $\varepsilon_{it} \sim N(0,\sigma_{it}^2)$, for $i=1:q$. The key technical benefit is $\mathrm{Cov}(\varepsilon_{it},\varepsilon_{js}) = 0$, for $i\neq j$ as well as for all $t,s$. Under conditionally independent priors over the coefficient processes and parameters, the model structure enables us to estimate $q$ univariate dynamic regression models separately, and in parallel. Gains in computational efficiency are relevant, in particular as $q$ increases, i.e., in higher-dimensional problems.

Then, posterior estimates from the decoupling step are fed into the recoupling step for forecasting and decisions. The recoupled model is basically based on eqn.~(\ref{eq:model}), where the $\A_t$ elements link (``cross-talk'') contemporaneous relationships among the $y_{it}$. Sequential forecasting and intervention analyses are straightforward with the reduced form equations.


\section{Contemporaneous relationship}

As discussed by the author in the paper, the ordering of the responses in $\y_t$, and more generally, the structure of $\A_t$ can be the issue. As far as an interest is forecasting is concerned, ordering is almost irrelevant because a predictive distribution relies only on the resulting covariance matrix $\bSigma_t$ in eqn.~ (\ref{eq:model}). However, some other analysis such as intervention and impulse response analysis may suffer from the issue.

There are mainly three formal approaches to addressing the structure of $\A_t$. One way is a use of economic theory or ``prior'' based on economic reasonings. In macroeconomics, the Cholesky-style decomposition has been widely used with the ordering determined based on some economic reasoning~\citep{Sims80}. For example, the interest rate is often placed last in the ordering as changes in the interest rate reflect contemporaneous changes in other macroeconomic variables such as output and inflation rate. \cite{Christianoetal99} propose a block recursive approach that restricts several elements in the triangular part to be zero.

The second approach is based on model fit and forecasting performance: one example is described in the SGDLM application (in Section 4.6). This gives an ``optimal'' pattern of the simultaneous parents in terms of forecasting, while some priors or constraints may be required if $q$ is quite large. The example in the paper sets $|pa(j)|=20$, for $q=401$, assuming relatively few series have conditional contemporaneous relationships with others. The third approach is a full analysis, searching for the best patterns of the simultaneous parents over all the possible combinations. When $q$ is small, it is possible to implement even if $\A_t$ is time-varying~\citep[see e.g.,][]{NakajimaWest13,NakajimaWest15}. However, if $q$ is large, it would be almost infeasible due to the computational burden in practice. Finally, a mixture of the theory-based approach and more data-based approaches could be suitable depending on data and context.




\newpage\setcounter{page}{1}\thispagestyle{empty}
\begin{center} 
{\LARGE\bf\blu Reply to Discussion of \\ ``Bayesian Forecasting of Multivariate Time Series: 
\smallskip

Scalability, Structure Uncertainty and Decisions''}
  
\bigskip\bigskip
{\Large  Mike West}\\

\bigskip
\today
\end{center} 

\section*{}
I am most grateful to the invited discussants,  Professor Chris Glynn and Dr.~Jouchi Nakajima, for their thoughtful and constructive comments and questions. 
Their discussion contributions speak clearly to some of the key areas of advance in Bayesian forecasting and time series modeling reviewed in the paper, and  critically address important areas of \lq\lq Challenges and Opportunities'' with some new suggestions and connections.  My responses here speak directly to their specific comments and questions.  I hope and expect that this conversation will additionally contribute to promoting new research developments in dynamic models for increasingly complex and challenging problems in multivariate time series analysis and forecasting-- and the broader fields of statistical modeling and decision analysis-- in the Akaike tradition.  

The discussants focus primarily on issues of model structure specification and learning in dynamic graphical models. These issues raise hard questions in multivariate models generally, as discussed in Section 2 of the paper. More specifically, they represent key current challenges in parental set specifications and modeling choices in  DDNMs (Section 3) and the more general class of SGDLMs (Section 4).   In two recent and current applied projects of my own and with collaborators, the exploration of multiple models based on ranges of parental sets has been-- and is--  the main effort in the  research enterprise. Some of the examples in the paper highlight these kinds of endeavors, using both traditional Bayesian model uncertainty approaches and shotgun stochastic search methods, while comparing  models on ranges of forecast and decision criteria as well as standard model probabilities. The model classes are now well understood, with immense flexibility to adapt to complex but inherently structured inter-dependencies among time series, and their changes in time. However, model choice and specification is challenging. 

Dr.~Nakajima  highlights the general problem based on his expertise and detailed experience with DDNMs linked, primarily but not exclusively, to macroeconomic time series modeling and forecasting.  With a cogent discussion some of the seminal background of decouple/recouple thinking in VAR and TV-VAR models, he wisely suggests that a blend of informed prior structuring coupled with empirical statistical model assessments is likely needed in applications in any other than a few dimensions. I very much agree. From the viewpoints of applied macroeconomics in areas such as monetary policy, bringing clear thinking about context and theoretically justified or required constraints on models is vital.  Then coupling that with (also clearly thought out) statistical assessments and comparisons is critical.  Traditionally, those applied areas have been dominated by the \lq\lq theory first'' view, but are increasingly integrating with the \lq\lq let the data speak'' view. Of course, the increasing impact of Bayesian methodology,  pioneered by influential time series econometricians~\citep[e.g.][]{Sims2012}, is central to this evolution, and TV-VAR models are routinely adopted.  My hope and expectation is that this will continue and that the approaches reviewed in my paper-- that extend traditional models with graphical/sparse structures more aggressively-- will be increasingly adopted in macroeconomics and in other fields.   That said, it remains the case that detailed evaluation of potentially many model choices-- testing partial constraints inspired by theory and context, and balanced by empirical testing with specific sets of defined forecast and/or decision goals in the use of the models-- will remain central to application. 

These challenges of model comparison with respect to parental set selection and structure are echoed in Professor~Glynn's comments and questions. Professor Glynn appropriately connects with more traditional Bayesian sparsity prior modeling approaches, whether \lq\lq point-mass mixtures'' or \lq\lq spike-and-slab'' structures.  I do agree that there are benefits of the latter over the former in technical senses, and some of the recent literature on bringing these ideas more aggressively into the sequential forward/filtering analysis of dynamic models is indeed interesting and exciting. The initial motivations for dynamic latent threshold models (LTMs, as discussed and noted in the several Nakajima {\em et al} references in the paper, and others) was in fact based on that traditional Bayesian thinking.  LTMs very naturally represent not only the interest in learning about changes over time in relevant variables-- here, relevant members of parental sets-- but also, in fact, imply \lq\lq smooth'' thresholding that corresponds to a class of dynamic spike-and-slab structures.   The applied relevance and impact of this thinking is, I believe, very clear from the several publications referenced in the paper, and other substantive applications such as~\cite{NakajimaEtAl2016}.   However, LTMs-- like other dynamic sparsity modeling approaches-- are inherently challenging to fit in a forward/sequential format, and some of the recent innovations in \lq\lq dynamic sparsity'' research that might be more conducive to efficient, and effective, sequential analysis are clearly of interest.  Bayesian optimization-based analysis as well as opportunities to exploit importance sampling in new ways are certainly promising directions, in my view.  In addition to the new directions by Rockova and McAlinn referenced by Professor Glynn,   I  note related developments
of~\cite{BittoFruhwirthSchnatter2019}, and quite novel dynamic sparsity structures-- and their Bayesian analyses-- of \cite{Irie2019} that open up new ground entirely with impressive examples. 

To both discussants and readers, however, I will summarize main concerns raised in the paper that are directly relevant to this core issue of model structure assessment. 

First, and critically: as dimensions scale the issues of predictor inter-dependencies generate messy problems of multiplicities leading-- inevitably-- to model uncertainties spread over increasing numbers of models 
that are exchangeable in any  practical sense. Typically, many parental set choices will generate similar \lq\lq fit'' to the data measured in  the usual ways and in terms of other specified forecast and decision outcomes.   Standard statistical thinking   fails as many   similar models are aggregated or selected, and the basic premise of sparsity modeling is violated~\citep[e.g.][]{GiannoneEtAlillusion2018}.
I encourage a more decision analytic view, i.e., selecting one or a small number of models, rather than the usual model averaging view. This, of course, requires articulation of forecasting and decision goals, and of relevant utility functions.  

Second,   to emphasize:  we model for reasons. Purely statistical assessments via posterior distributions over  models-- whether combined with insightful theoretical constraints or not-- are valid only if we choose \lq\lq purely statistical'' to define utility functions for model uses.  Examples in the paper  highlight this, and I hope that this paper and discussion will aid the broader community in considering modeling usage goals as part of the broader enterprise in model comparison and selection.  

Third, but not at all least: dynamics and sequential settings.  Much traditional Bayesian machinery-- dominated by MCMC in the last three decades-- simply does not translate to the sequential  setting.     Currently fashionable methods of sequential Monte Carlo face tremendous challenges in any but small problems, and have yet to properly impact in large-scale applications. New ideas and methodology for finding, evaluating, comparing and combining models-- generally as well as in connection with parental sets in DDNMs and SGDLMs-- are critically needed in the sequential context.   Some of the perspective mooted in Section 4 of the paper-- of integrating more formal Bayesian decision theoretic thinking into the model uncertainty context-- seem very worth embracing and developing.  The conceptual advances in~\cite{LavineLindonWest2019avs} represent some of my own recent thinking and collaborative development in this direction. Adopting such perspectives will, I predict,  open up opportunities for core research and advance methodology in the Akaike spirit: challenging statistical modeling and decision analysis issues motivated by hard, important applications, that engage existing and new researchers in conceptual and theoretical innovation to bring back to address  those real-world problems.

\bigskip

\end{document}